\documentclass[12pt]{iopart}
\usepackage{iopams}
\usepackage{amsfonts}
\usepackage{amsthm}
\usepackage{amstext}
\usepackage{amssymb}
\usepackage{amsbsy}
\usepackage{amscd}
\usepackage{latexsym}
\usepackage[dvips]{graphics}
\usepackage[pdf]{graphicx}
\usepackage{a4wide}
\usepackage{geometry}
\usepackage{epsfig}
\usepackage{fancyhdr}
\usepackage{color} 
\usepackage{pstricks}
\usepackage{indentfirst}
\usepackage[tight]{subfigure}
\usepackage{psfrag}
\usepackage{ifthen} 
\usepackage{eso-pic}
\input amssym.def
\input amssym.tex
\eqnobysec
\makeatletter
\def\Ddots{\mathinner{\mkern1mu\raise\p@
 \vbox{\kern7\p@\hbox{.}}\mkern2mu
 \raise4\p@\hbox{.}\mkern2mu\raise7\p@\hbox{.}\mkern1mu}}
\makeatother

\let\oldsqrt\sqrt
\def\sqrt{\mathpalette\DHLhksqrt}
\def\DHLhksqrt#1#2{%
\setbox0=\hbox{$#1\oldsqrt{#2\,}$}\dimen0=\ht0
\advance\dimen0-0.2\ht0
\setbox2=\hbox{\vrule height\ht0 depth -\dimen0}%
{\box0\lower0.4pt\box2}}

\begin{document}
\title{Spatial chaos of an extensible conducting rod in a uniform magnetic field}
\author{D~Sinden and G~H~M~van~der~Heijden}
\address{Centre for Nonlinear Dynamics, University College London, Chadwick Building, Gower Street, London WC1E 6BT, UK} 
\eads{\mailto{d.sinden@ucl.ac.uk, g.heijden@ucl.ac.uk}}
\begin{abstract}
The equilibrium equations for the isotropic Kirchhoff rod are known to form an integrable system. It is also known that the effects of extensibility and shearability of the rod do not break the integrable structure. Nor, as we have shown in a previous paper does the effect of a magnetic field on a conducting rod. Here we show, by means of Mel'nikov analysis, that, remarkably, the combined effects do destroy integrability; that is, the governing equations for an extensible current-carrying rod in a uniform magnetic field are nonintegrable. This result has implications for possible configurations of electrodynamic space tethers and may be relevant for electromechanical devices.
\end{abstract}
\newtheorem{thm}{Theorem}[section]
\newtheorem{conj}[thm]{Conjecture}
\newtheorem{cor}[thm]{Corollary}
\newtheorem{lem}[thm]{Lemma}
\newtheorem{prop}[thm]{Proposition}
\newtheorem{rem}[thm]{Remark}
\newtheorem{defin}[thm]{Definition}
\newtheorem{cla}[thm]{Claim}
\newtheorem{example}[thm]{Example}

\section{Introduction} \label{sec:intro}

The geometrically exact static equilibrium equations for a uniform symmetric
(i.e., transversely isotropic) elastic rod are well-known to
be completely integrable \cite{Kehrbaum97b}. In fact, there is a close relationship between these equations and those describing the dynamics of spinning tops. It is also known that some perturbations of the rod equations are integrable, but
that others are not. For instance, anisotropy of the cross-section
\cite{Mielke88} and intrinsic curvature \cite{Champneys97a} destroy integrability, as does the effect of gravity \cite{Gottlieb99}, but extensibility and shearability \cite{Stump00} do not, nor does the effect of an external force due to a uniform magnetic field \cite{Sinden08}.

In this paper we show by means of a perturbation analysis that the combined effect of extensibility and magnetic field, remarkably, leads to nonintegrability, even though each of these effects individually would not destroy integrability. The results may be relevant for the study of (localised) spatial configurations of
electrodynamic space tethers, i.e., conducting cables that exploit the
earth's magnetic field to generate thrust and drag (Lorentz) forces for
manoeuvring \cite{Cartmell08,Valverde08}.

The perturbation theory we use is the Hamiltonian version of Mel'nikov theory as developed by Holmes \& Marsden \cite{Holmes83}. This theory considers the break-up of a homoclinic orbit of the unperturbed integrable system by detecting transverse intersections of stable and unstable manifolds of a perturbed saddle-type solution. Specifically, the so-called Mel'nikov integral measures the distance between stable and unstable manifolds. Simple zeroes of this integral correspond to transverse intersections which imply
complicated horseshoe dynamics and hence nonintegrability.

From standard results of dynamical systems theory one would also generically
expect in this situation the existence of a multiplicity of multipulse
homoclinic orbits \cite{Belyakov90}, corresponding to arbitrary many localised solutions of the rod, and we show numerically that this is indeed the case.

Mel'nikov analyses have been applied before to prove nonintegrability of
anisotropic rods \cite{Mielke88}, intrinsically curved rods \cite{Leung04} and heavy rods \cite{Gottlieb99}. The first two of these studies employed a formulation in terms of Deprit-Andoyer variables. Here we use a formulation in terms of the more common Euler angles. Multimodal configurations for anisotropic and intrinsically curved rods have been investigated numerically in \cite{Champneys96b,Heijden98a} and \cite{Champneys97a}, respectively.

To apply perturbation theory in the case that two effects are simultaneously present one has to make assumptions on the relative scale of the two effects. Since Mel'nikov analysis requires closed-form expressions for the homoclinic orbit and these expressions are more readily obtained for the extensible rod than for the magnetic rod, the Mel'nikov theory is applied with the magnetic field as a perturbation of the extensible rod. However, we present numerical evidence, in the form of chaotic Poincar\'e sections, that suggests that integrability is also broken in the opposite scaling, i.e., if extensibility is viewed as a perturbation of the magnetic rod.

The structure of the paper is as follows. In Section \ref{sec:model} the governing equations are presented as a non-canonical Hamiltonian system. In Section \ref{sec:melnikov_thm} Hamiltonian Mel'nikov theory is briefly reviewed. In Section \ref{sec:reduction} the system of equations is reduced to a canonical system by using the Casimirs of the Poisson bracket. In Section \ref{subsec:extensible} the homoclinic orbits of the unperturbed system are calculated, after which the Mel'nikov analysis is performed in Section \ref{sec:melnikov}. Multimodal homoclinic orbits and fractal Poincar\'e plots, the signatures of spatial chaos, are computed in Section \ref{sec:numerics} and shown to persist in regions of the parameter space well away from the asymptotic region where the Mel'nikov result is valid. Section \ref{sec:conc} closes this study with some concluding remarks.

\section{The Cosserat Theory of Elastic Rods} \label{sec:model} %
\subsection{Kinematic equations} \label{subsec:kinematic}
In Cosserat theory a rod is characterised by a space curve $\boldsymbol{r}\left(s\right)$, describing the centreline of the rod, and an attached right-handed orthonormal triad of directors 
${ \left\{\boldsymbol{d}_{1}(s), \boldsymbol{d}_{2}(s), \boldsymbol{d}_{3}(s) \right\} }$, describing the varying orientation of the cross-section \cite{Antman05}. Here $s$ is an arbitrary parameter.

On introducing a right-handed orthonormal frame
${ \left\{ \boldsymbol{e}_{1}, \boldsymbol{e}_{2}, \boldsymbol{e}_{3} \right\} }$ fixed in space we can write
\begin{eqnarray}
\boldsymbol{d}_{i} & = {R}\boldsymbol{e}_{i},
\label{eq:frame}
\end{eqnarray}
where $R$ is a rotation matrix, i.e., an element of the Lie group SO(3). It is convenient to introduce the `hat map' isomorphism of the corresponding Lie algebra $\mathfrak{so}\left(3\right)$:
\begin{equation}
\mathbb{R}^{3} \longrightarrow \mathfrak{so}\left(3\right): \quad\quad
a = \left( a_{1}, a_{2}, a_{3} \right) \mapsto \hat{a} = 
\left( \begin{array}{ccc}
0 & -a_{3} & a_{2} \\
a_{3} & 0 & -a_{1} \\
-a_{2} & a_{1} & 0 
\end{array} \right),
\label{eq:isomorphism}
\end{equation}
so that ${\hat{a}b = a \times b}$. Differentiating \eref{eq:frame} then gives
\begin{eqnarray}
\boldsymbol{d}^{\prime}_{i} & = {R}^{\prime}_{} \boldsymbol{e}_{i}^{} = {R}^{\prime} {R}^{T} \boldsymbol{d}_{i} = :  \hat{\boldsymbol{u}} \boldsymbol{d}_{i} = \boldsymbol{u} \times \boldsymbol{d}_{i},
\label{eq:directors}
\end{eqnarray}
where the prime denotes differentiation with respect to $s$ and $\boldsymbol{u}$ is the vector of generalised strains associated with bending and twisting. The body components $u_i=\boldsymbol{u}\cdot\boldsymbol{d}_i$ are the curvatures ($i = 1, 2$) and the twist ($i = 3$) of the rod, for which, from \eref{eq:directors}, we can write
\begin{eqnarray}
u_{i}^{} & = \frac{1}{2} \varepsilon_{ijk}^{} \boldsymbol{d}^{\prime}_{j} \cdot \boldsymbol{d}_{k}^{}. \label{eq:curvatures}
\end{eqnarray}
The second vector of strains $\boldsymbol{v}$ is given by
\begin{eqnarray}
\boldsymbol{r}^{\prime} & = \boldsymbol{v}. \label{eq:centreline}
\end{eqnarray}
The body components ${ v_{1}=\boldsymbol{v}\cdot\boldsymbol{d}_{1} }$, ${v_{2}=\boldsymbol{v}\cdot\boldsymbol{d}_{2} }$ and ${v_{3}=\boldsymbol{v}\cdot\boldsymbol{d}_{3} }$ are the strains associated with stretching and shear. For an unshearable rod we have $v_1=0=v_2$. For an inextensible rod we have ${\left| \boldsymbol{r}^{\prime} \right| = 1}$. The strain $v_3$ actually represents the ratio of deformed to reference volume. Since a real rod cannot be compressed to a point, it is natural to impose the condition $v_3>0$.
For an inextensible and unshearable rod the centreline equation \eref{eq:centreline} becomes
\begin{eqnarray}
\boldsymbol{r}^{\prime} & = \boldsymbol{d}_{3} \label{eq:inextensible}
\end{eqnarray}
and the parameter $s$ can be interpreted as the arclength of the rod.

It will be convenient in the following sections to express components of vectors with respect to the director (or body) frame; for any vector $\boldsymbol{p}$ the triple of components
$\left( \boldsymbol{p}\cdot \boldsymbol{d}_{1}, \boldsymbol{p}\cdot \boldsymbol{d}_{2}, \boldsymbol{p}\cdot \boldsymbol{d}_{3} \right) $
will be denoted by the sans-serif symbol $\mathsf{p}$.

\subsection{Constitutive relations} \label{subsec:constitutive}

We assume the rod to be hyperelastic, i.e., we assume that there is a strain energy density function
$\mathcal{W}=\mathcal{W}\left(\mathsf{u}-\mathsf{u}_{0},\mathsf{v}-\mathsf{v}_{0},s \right)$ such that the components of the force $\mathsf{n}=\left(n_{1},n_{2},n_{3}\right)$ and moment $\mathsf{m}=\left(m_{1},m_{2},m_{3}\right)$ in the body are given by
\begin{equation}
m_{i} = \frac{\partial \mathcal{W}}{\partial u_{i}} \quad \mbox{and} \quad n_{i} = \frac{\partial \mathcal{W}}{\partial v_{i}}. \label{eq:constit}
\end{equation}
Here $\mathsf{u}_{0}$ and $\mathsf{v}_{0}$ describe the configuration of the unstressed rod.

We will consider the important special case, often called linearly elastic, where the strain energy is quadratic in the strains:
\begin{eqnarray}
\mathcal{W} \left( {\mathsf{u}}, {\mathsf{v}} \right) & = \frac{1}{2}B^{}_{1}u_{1}^{2} + \frac{1}{2}B^{}_{2}u_{2}^{2} + \frac{1}{2} C u_{3}^{2} + \frac{1}{2} H v_{1}^{2} + \frac{1}{2} J {v_{2}^{2}} + \frac{1}{2} K \left(v_{3}-1\right)^{2}, \label{eq:hyperelastic}
\end{eqnarray}
where $B_{1}$ and $B_{2}$ are the principal bending stiffnesses about $\boldsymbol{d}_{1}$ and $\boldsymbol{d}_{2}$, respectively, and $C$ is the torsional stiffness about $\boldsymbol{d}_{3}$. The constants $H$ and $J$ are the transverse shear stiffnesses and $K$ is the axial stiffness. In the case of an isotropic rod, ${B_{1}=B_{2}=:B}$ and ${H=J}$.

\subsection{Equilibrium equations} \label{subsec:equilibrium}

The equilibrium equations for the internal force $\boldsymbol{n}$ and moment $\boldsymbol{m}$ in an elastic rod are \cite{Antman05}
\begin{eqnarray}
&& \boldsymbol{n}^{\prime} + \boldsymbol{f} = \boldsymbol{0}, \label{eq:force} \\
&& \boldsymbol{m}^{\prime} + \boldsymbol{r}^{\prime} \times \boldsymbol{n} + \boldsymbol{l} = \boldsymbol{0}, \label{eq:moment}
\end{eqnarray}
where we have allowed for external distributed loads $\boldsymbol{f}$ and $\boldsymbol{l}$. The only distributed load we shall consider is that due to a magnetic field, in which case $\boldsymbol{f}$ is given by the Lorentz force, while 
$\boldsymbol{l}=\boldsymbol{0}$. Assuming the rod to be conducting and to carry a current ${\boldsymbol{I}=I\boldsymbol{r}^{\prime}}$ the Lorentz force experienced when placed in a magnetic field $\bar{\boldsymbol{B}}$ is
\begin{equation}
\boldsymbol{f}=\boldsymbol{I}\times\bar{\boldsymbol{B}} = I\boldsymbol{r}^{\prime}\times\bar{\boldsymbol{B}}. \label{eq:lorentz}
\end{equation}
We assume both current and magnetic field to be uniform and align the $\boldsymbol{e}_3$ vector of the fixed frame with the field, so that $\bar{\boldsymbol{B}}=\bar{B}\boldsymbol{e}_3$. Let $\lambda=I\bar{B}$, then the equilibrium equations when written in the director frame take the form of a non-canonical Hamiltonian system:
\begin{eqnarray}
\fl
\left(
\begin{array}{c}
\mathsf{m} \\
\mathsf{n} \\
\mathsf{e}_{3}
\end{array}
\right)^{\prime} & =
\mathcal{J} \left({\mathsf{m}},{\mathsf{n}},\mathsf{e}_{3}; \lambda \right) \nabla \mathcal{H}\left({\mathsf{m}},{\mathsf{n}} \right) \quad \mbox{with} \quad \mathcal{J} = -\mathcal{J}^{T} & = \left( 
\begin{array}{ccc}
\hat{\mathsf{m}} & \hat{\mathsf{n}} & \hat{\mathsf{e}_{3}} \\
\hat{\mathsf{n}} & \lambda\hat{\mathsf{e}_{3}} & \mathsf{0} \\
\hat{\mathsf{e}_{3}} & \mathsf{0} & {\mathsf{0}}  
\end{array}
\right) \label{eq:noncanonical_structure}
\end{eqnarray}
with Hamiltonian
\begin{equation}
\mathcal{H}\left(\mathsf{m},\mathsf{n}\right) = \frac{1}{2} \mathsf{m} \cdot \mathsf{u(\mathsf{m})} + \frac{1}{2}\mathsf{n}\cdot\left(\mathsf{v}(\mathsf{n})-\mathsf{d}_{3}\right) + \mathsf{d}_{3}\cdot\mathsf{n}, \label{eq:magnetic_ham}
\end{equation}
where $\mathsf{d}_{3}=\left(0,0,1\right)$.
In writing down this equation we have used \eref{eq:directors}, \eref{eq:centreline}
and the hat isomorphism~\eref{eq:isomorphism}. This formulation follows that of \cite{Sinden08} but now allows for the effects of extensibility and shearability.
As in the inextensible/unshearable case, the Hamiltonian is the same as that of the non-magnetic rod (cf.~\cite{Dichmann96c}): the effect of the magnetic field is only present in the structure matrix $\mathcal{J}$.

By introducing the Poisson bracket
\begin{eqnarray}
\fl
\left\{ f,g \right\}_{\left({\mathsf{m}},{\mathsf{n}}, {\mathsf{e}_{3}} \right)} & =
- {\mathsf{m}}\cdot\left(\nabla_{{\mathsf{m}}}f \times \nabla_{{\mathsf{m}}}g \right) 
- {\mathsf{n}}\cdot\left(\nabla_{{\mathsf{m}}}f \times \nabla_{{\mathsf{n}}}g + \nabla_{{\mathsf{n}}}f \times \nabla_{{\mathsf{m}}}g\right) \nonumber \\
\fl
& \hspace{1.25cm} - \underbrace{{\mathsf{e}_{3}} \cdot \left(\nabla_{\mathsf{m}}f \times \nabla_{\mathsf{e}_{3}}g +
\nabla_{\mathsf{e}_{3}}f \times \nabla_{{\mathsf{m}}}g \right)}_{\mbox{evolution of field}}
- \underbrace{\lambda \mathsf{e}_{3} \cdot \left( \nabla_{\mathsf{n}}f \times \nabla_{\mathsf{n}}g \right)}_{\mbox{effect of field}},
\label{eq:magnetic_bracket}
\end{eqnarray}
on ${ \left( \mathsf{m}, \mathsf{n}, \mathsf{e}_{3} \right) }$ the equilibrium equations can also be written as
\begin{eqnarray}
\mathsf{m}^{\prime} & = \left\{\mathsf{m}, \mathcal{H} \right\}_{\left({\mathsf{m}},{\mathsf{n}}, {\mathsf{e}_{3}} \right)} = \mathsf{m}\times\mathsf{u} + \mathsf{n}\times\left(\mathsf{d}_{3}+\mathsf{v}\right), \label{eq:magnetic_moment} \\
\mathsf{n}^{\prime} & = \left\{\mathsf{n}, \mathcal{H} \right\}_{\left({\mathsf{m}},{\mathsf{n}}, {\mathsf{e}_{3}} \right)}  \,\, = \mathsf{n}\times\mathsf{u} + \lambda\mathsf{e}_{3}\times\left(\mathsf{d}_{3}+\mathsf{v}\right), \label{eq:magnetic_force} \\
\mathsf{e}_{3}^{\prime} & = \left\{\mathsf{e}_{3}, \mathcal{H} \right\}_{\left({\mathsf{m}},{\mathsf{n}}, {\mathsf{e}_{3}} \right)} = \mathsf{e}_{3}\times\mathsf{u}. \label{eq:magnetic_vector}
\end{eqnarray}
Note that the Poisson bracket is an extension of the usual Kirchhoff bracket by two terms. The first is a semidirect term \cite{Holm98} describing the evolution of the magnetic field in the director frame (this term does not affect the force and moment balance since the Hamiltonian is independent of $\mathsf{e}_{3}$). The second is a cocycle known as a Leibniz extension~\cite{Thiffeault01} describing the effect of the magnetic field on the conducting rod. Note that \eref{eq:magnetic_vector} is just the equation $\boldsymbol{e}_3^{\prime}=\boldsymbol{0}$ written in the director frame.

Note on notation: an ornamented gradient symbol is used for gradients with respect to the indicated fields, while the unornamented symbol will always denote the gradient with respect to all three fields ${ \left( \mathsf{m}, \mathsf{n}, \mathsf{e}_{3} \right) }$.

The non-canonical system \eref{eq:magnetic_moment}--\eref{eq:magnetic_vector} has three Casimirs, given by
\begin{eqnarray}
C_{1} & = \frac{1}{2}\mathsf{n} \cdot \mathsf{n} + \lambda \mathsf{m} \cdot \mathsf{e}_{3}, \label{eq:magnetic_casimirs1} \\
C_{2} & = \mathsf{e}_{3} \cdot \mathsf{n}, \label{eq:magnetic_casimirs2} \\
C_{3} & = \mathsf{e}_{3} \cdot \mathsf{e}_{3}. \label{eq:magnetic_casimirs3}
\end{eqnarray} 
The magnitude of force is not conserved if a magnetic field is present, but as a result of rotational symmetry the force component in the direction of the field is conserved. Casimir~\eref{eq:magnetic_casimirs1} does not seem to have a physical interpretation. Naturally, $C_3=1$.

In the special case of inextensible/unshearable and isotropic rods two first integrals emerge:
\begin{eqnarray}
I_{1} & = B\mathsf{m}\cdot\mathsf{d}_3 \quad\quad (\mbox{if} \quad {J=H} \quad \mbox{and} \quad {B_1=B_2=B}), \label{eq:magnetic_lagrange} \\
I_{2} & = \mathsf{n}\cdot\mathsf{m} + B\lambda \mathsf{e}_{3} \cdot \mathsf{d}_{3} \quad\quad (\mbox{if} \quad {J=H=K=0} \quad \mbox{and} \quad {B_1=B_2=B}). \label{eq:int2}
\end{eqnarray}
As in the Kirchhoff case, the first of these expresses conservation of twist in the rod. The second integral does not seem to have a physical interpretation, but in the limit $\lambda\to 0$ it reduces to the familiar conservation of torque about the loading axis. It is a straightforward task to check that all the first integrals~\eref{eq:magnetic_ham},~\eref{eq:magnetic_lagrange} and~\eref{eq:int2} are independent and in involution with respect to the Poisson bracket~\eref{eq:magnetic_bracket}, and therefore the system is completely integrable.

Note that the twist integral $I_1$ only requires isotropy, while the integral $I_2$ requires isotropy and inextensibility/unshearability. Indeed, it is the main purpose of this paper to show that the combined effect of inextensibility/unshearability and magnetic field leads to nonintegrability.
For this we use Mel'nikov theory, which we review next.

\section{Mel'nikov theory} \label{sec:melnikov_thm}

For Hamiltonian systems Mel'nikov's original perturbation analysis needs to be adapted. We use the results for two-degrees-of-freedom systems presented in \cite{Holmes83,Mielke88}.

Consider a Hamiltonian that depends on a small parameter $\epsilon$ in the form
\begin{equation}
\mathcal{H}_{}\left(q,p,\varphi,I\right) = \mathcal{H}_{0}\left(q,p,I\right) + \epsilon \mathcal{H}_{1}\left(q,p,\varphi,I\right) + \mathcal{O}\left( \epsilon^{2}\right), \label{eq:perturbed_hamiltonian}
\end{equation}
where $(q,p)$ are conjugate variables and $(I,\varphi)$ are action-angle variables
such that $\mathcal{H}_1$ is $2\pi$-periodic in $\varphi$. For $\epsilon=0$ $\varphi$ is a cyclic variable (hence $I$ is a first integral) and Hamilton's equations are completely integrable. We assume that the unperturbed Hamiltonian $\mathcal{H}_{0}$ satisfies the following two conditions:
\begin{itemize}
\item[(i)] For some $I=I_0$ Hamilton's equations corresponding to $\mathcal{H}_{0}$ possess a homoclinic orbit ${ \left( \bar{q}\left(t\right), \bar{p}\left(t\right) \right) }$ to a hyperbolic fixed point ${\left( q_0, p_0 \right) }$ at Hamiltonian level $h=\mathcal{H}_{0}(\bar{q},\bar{p},I_0)$.
\item[(ii)] The frequency
\begin{eqnarray}
\omega_{0} & := \frac{\partial \mathcal{H}_{0}}{\partial I}(\bar{q},\bar{p},I_0) \label{eq:Omega_0} 
\end{eqnarray}
of the unperturbed system satisfies ${\left| \omega_{0} \right| \ge \nu > 0}$ for some ${\nu \in \mathbb{R}}$ and ${\forall t \in \left(-\infty,+\infty\right)}$.
This condition means that $\varphi$ is a time-like variable and allows the unperturbed system to be reduced to the $(q,p)$ space with $\varphi$ as the independent variable.
\end{itemize}

Now define the Mel'nikov function
\begin{eqnarray}
\mathcal{M}\left(\varphi_{0}\right) & = \int^{+\infty}_{-\infty} \left\{ \mathcal{H}_{0}, \frac{\mathcal{H}_{1}}{\omega_{0}} \right\}_{\left(q,p\right)} \,\mathrm{d}t,
\label{eq:melnikov_bracket}
\end{eqnarray}
where the canonical Poisson bracket $\{f,g\}_{(q,p)}=\frac{\partial f}{\partial q}\frac{\partial g}{\partial p}-\frac{\partial f}{\partial p}\frac{\partial g}{\partial q}$ is evaluated at the homoclinic orbit and $\varphi(t)=\int_0^t\omega_0(\bar t)\,\mathrm{d}\bar t+\varphi_0$.
We then have the following result:
\begin{thm}\label{thm:melnikov}
For $\epsilon\neq 0$ sufficiently small, if $\mathcal{M}\left(\varphi_{0}\right)$ has simple zeroes $\bar{\varphi}_0$, that is,
\begin{equation}
\mathcal{M}\left(\bar{\varphi}_{0}^{}\right)=0 \quad \mbox{and} \quad \frac{\partial \mathcal{M}}{\partial \varphi_{0}^{}}(\bar{\varphi}_0) \ne 0,
\end{equation}
then the stable and unstable manifolds of the perturbed hyperbolic invariant set (a periodic solution in the four-dimensional system) intersect transversally. If, on the other hand, $\mathcal{M}\left(\varphi_{0}\right)$ is bounded away from zero, then the manifolds do not intersect.
\end{thm}
\begin{proof}[Proof of~\ref{thm:melnikov}]
See~\cite{Mielke88,Holmes83,Guckenheimer83}.
\end{proof}
The existence of transverse homoclinic orbits implies that the `dynamics' near the hyperbolic saddle is `chaotic' in the sense that the following holds:
\begin{cor} \label{cor:chaos}
The Poincar\'e map associated with $\mathcal{H}$ on the homoclinic level set $\mathcal{H}^{-1}(h)$ has a hyperbolic, non-wandering Cantor set on which the map is conjugate to a Bernoulli shift of finite type.
\end{cor}
\begin{proof}[Proof of~\ref{cor:chaos}]
See~\cite{Mielke88} and \cite{Guckenheimer83}.
\end{proof}
This in turn implies
\begin{cor} \label{cor:melnikov}
The Hamiltonian $\mathcal{H}$ has no analytic conserved quantities independent of $\mathcal{H}$ itself, i.e., the corresponding Hamiltonian system is nonintegrable.
\end{cor}
\begin{proof}[Proof of~\ref{cor:melnikov}]
See~\cite{Mielke88}.
\end{proof}

\section{Reduction of the magnetic rod equations to a canonical system} \label{sec:reduction}

In this section the three Casimirs~\eref{eq:magnetic_casimirs1}--\eref{eq:magnetic_casimirs3} are used to reduce the nine-dimensional non-canonical Hamiltonian system~\eref{eq:magnetic_moment}--\eref{eq:magnetic_vector} in ${\left(\mathsf{m},\mathsf{n},\mathsf{e}_{3}\right)}$ to a six-dimensional canonical Hamiltonian system in terms of Euler angles and their canonical momenta $(q,p)={\left(\theta,\psi,\phi,p_\theta,p_\psi,p_\phi\right)}$. The reduction follows~\cite{Sinden08} but now allows for extensibility and shearability of the rod. The reduction in \cite{Sinden08} was shown to be canonical on the condition that the force $\boldsymbol{n}$ and magnetic field $\bar{\boldsymbol{B}}$
are not aligned. This result trivially extends to the present case as the effects of inextensibility and shearability do not enter the structure matrix, only the Hamiltonian.
\par
Let
\begin{eqnarray*}
\fl
R & = 
\left(
\begin{array}{ccc}
\cos\theta\cos\phi\cos\psi-\sin\phi\sin\psi & \cos\theta\cos\phi\sin\psi+\cos\psi\sin\phi & -\sin\theta\cos\phi \\
-\cos\theta\sin\phi\cos\psi-\cos\phi\sin\psi & -\cos\theta\sin\phi\sin\psi+\cos\phi\cos\psi & \sin\theta\sin\phi \\
\sin\theta\cos\psi & \sin\theta\sin\psi & \cos\theta
\end{array}
\right) \nonumber
\end{eqnarray*}
be a parametrisation of the rotation matrix $R$ in~\eref{eq:frame} in terms of Euler angles. Here $\theta$ is the angle the tangent to the rod makes with the magnetic field, $\psi$ is the azimuthal angle about the field direction and $\phi$ is the twist angle about the centreline of the rod. It follows that for the triple $\mathsf{e}_{3}$ we have
\begin{eqnarray}
\mathsf{e}_{3}\left(q\right) & = R\left(q\right) \mathsf{k} =
\left( -\sin\theta\cos\phi, \sin\theta\sin\phi, \cos\theta \right)^{T},
\label{eq:canonical_magnetic}
\end{eqnarray}
where $\mathsf{k}=\left(0,0,1\right)^T$. On inserting the Euler angles into the strains~\eref{eq:curvatures} and \eref{eq:centreline} and using the constitutive relations~\eref{eq:constit} the moments are found to be
\begin{eqnarray}
\mathsf{m} & =
\left(
\begin{array}{c}
m_1 \\
m_2 \\
m_3
\end{array}
\right)
=
\left( 
\begin{array}{c}
B_1 (\theta^{\prime} \sin\phi - \psi^{\prime} \sin\theta \cos\phi) \\
B_2 (\theta^{\prime} \cos\phi + \psi^{\prime} \sin\theta \sin\phi) \\
C (\phi^{\prime} + \psi^{\prime} \cos\theta)
\end{array}
\right)
=Lp,
\label{eq:canonical_moment}
\end{eqnarray}
where
\begin{eqnarray}
L & = \frac{1}{\sin\theta} 
\left( 
\begin{array}{ccc}
\sin\theta\sin\phi & -\cos\phi &  \cos\theta\cos\phi \\
\sin\theta\cos\phi &  \sin\phi & -\cos\theta\sin\phi \\
0 & 0 & \sin\theta 
\end{array}
\right)
\nonumber
\end{eqnarray}
and the canonical momenta defined by $p_\theta=\partial \overline \mathcal{W}(q,q')/\partial \theta'$, $p_\psi=\partial \overline \mathcal{W}(q,q')/\partial \psi'$, $p_\phi=\partial \overline \mathcal{W}(q,q')/\partial \phi'$, with $\overline \mathcal{W}(q,q')=\mathcal{W}(\mathsf{u}(q,q'))$ in terms of the strain energy function $\mathcal{W}$ defined in \eref{eq:hyperelastic}.

For the force we can write ${\mathsf{n}=R\left(q\right)w\left(q,p\right)}$, for some non-constant triple $w$. By decomposing $w$ into parts perpendicular and parallel to $\mathsf{k}$ and using the Casimirs~\eref{eq:magnetic_casimirs1} and~\eref{eq:magnetic_casimirs2} we obtain \cite{Sinden08}
\begin{eqnarray}
\fl
\mathsf{n} & = 
C_{2}^{}
\left( 
\begin{array}{c}
-\sin\theta\cos\phi \\
\sin\theta\sin\phi \\
\cos\theta
\end{array}
\right) + \sqrt{ 2 C_{1}^{} - C_{2}^{2} - 2 \lambda p_{\psi}^{}  }
\left(\begin{array}{c}
 \cos\theta\cos\phi\cos\psi-\sin\phi\sin\psi \\
-\cos\theta\sin\phi\cos\psi-\cos\phi\sin\psi \\
 \sin\theta\cos\psi
\end{array}
\right). \label{eq:canonical_force}
\end{eqnarray}

It was shown in \cite{Sinden08} that the transformation \eref{eq:canonical_magnetic}--\eref{eq:canonical_force} is canonical provided that 
the force and magnetic field (or $\boldsymbol{e}_3$) are not aligned, i.e.,
\begin{eqnarray}
2 C_{1}^{} - C_{2}^{2} - 2 \lambda p_{\psi} & \neq 0. \label{eq:alignment_condition}
\end{eqnarray}
(It was also shown that if \eref{eq:alignment_condition} holds anywhere along the rod it holds everywhere, and in that case all solutions are straight twisted rods, aligned with the magnetic field.) In the linearly elastic case the Hamiltonian \eref{eq:magnetic_ham} transforms into
\begin{equation}
\mathcal{H}\left( \theta, \psi, \phi, p_{\theta}, p_{\psi}, p_\phi \right)= \frac{m_1^2}{2B_1}+\frac{m_2^2}{2B_2}+\frac{m_3^2}{2C}+\frac{n_1^2}{2H}+\frac{n_2^2}{2J}+\frac{n_3^2}{2K}+n_3,
\label{eq:ham_canon} 
\end{equation}
with $m_i$, $n_i$ given in terms of the canonical variables by \eref{eq:canonical_moment} and \eref{eq:canonical_force}. In the isotropic case ($B_1=B_2=:B$, $H=J$) the Hamiltonian reduces further to
\begin{eqnarray}
\fl
\mathcal{H}\left( \theta, \psi, p_{\theta}, p_{\psi}, p_{\phi} \right) & = \frac{1}{2 B} p_{\theta}^{2} + \frac{1}{2 B} \left( \frac{p_{\psi}-p_{\phi}\cos\theta}{\sin\theta} \right)^{2} + C_{2}\cos\theta\left( \frac{C_{2}}{2}\left(\frac{1}{K}-\frac{1}{J}\right)\cos\theta + 1 \right) \nonumber \\
\fl
& \hspace{0.75cm} + \left( C_{2}\left(\frac{1}{K}-\frac{1}{J}\right)\cos\theta + 1 \right)\sin\theta\cos\psi\sqrt{ 2 C_{1}^{} - C_{2}^{2} - 2 \lambda p_{\psi} } \nonumber \\ 
\fl
& \hspace{1.25cm} + \frac{1}{2}\left(\frac{1}{K}-\frac{1}{J}\right)\sin^{2}\theta\cos^{2}\psi\left( 2 C_{1}^{} - C_{2}^{2} - 2 \lambda p_{\psi} \right) -\frac{\lambda}{J}p_\psi, \label{eq:full_ham}
\end{eqnarray}
where we have dropped the $p_\phi^2$ term, which is constant since $\phi$ is a cyclic variable: $p_{\phi}=m_3=I_1/B$. If, in addition, the rod is inextensible and unshearable ($H=J=K=0$) then Hamilton's equations corresponding to \eref{eq:full_ham} have $I_2$ in \eref{eq:int2} as a first integral, which in canonical variables takes the form
\begin{eqnarray}
\fl
I_{2} & = \lambda B \cos\theta + C_{2}^{} p_{\psi} - \sqrt{ 2 C_{1}^{} - C_{2}^{2} - 2 \lambda p_{\psi}^{} } \left( p_{\theta}\sin\psi - \cos\psi \left( \frac{p_{\phi} - p_{\psi}\cos\theta}{\sin\theta} \right) \right),
\label{eq:constraint}
\end{eqnarray}
rendering the system completely integrable.

Finally, we use the constants $C_2$ and $m_3=I_1/B=p_\phi$ to introduce dimensionless quantities by setting
\begin{eqnarray}
&& t = s\frac{m_3}{B}, \quad \bar{p}_\theta = \frac{p_\theta}{m_3}, \quad \bar{p}_\psi = \frac{p_\psi}{m_3} \quad \bar{\lambda}=\frac{\lambda m_3}{C_2^2}, \quad \mu=\frac{2C_1-C_2^2}{C_2^2}, \\
&& \gamma = C_2\left(\frac{1}{K}-\frac{1}{J}\right), \quad \delta = \frac{C_2}{J}, \quad m = \frac{m_3}{ \sqrt{B C_2} }, \label{eq:nondim_1}
\end{eqnarray}
so that the dimensionless Hamiltonian $\bar{\mathcal{H}}=\mathcal{H}B/m_3^2$ and integral $\bar{I}_2=I_2/(C_2m_3)$ become \\ \\
\begin{eqnarray}
\bar{\mathcal{H}}\left( \theta, \psi, \bar{p}_{\theta}, \bar{p}_{\psi}\right)
 = & \frac{1}{2} \bar{p}_{\theta}^{2} + \frac{1}{2} \left( \frac{\bar{p}_{\psi}-\cos\theta}{\sin\theta} \right)^{2} + \frac{\cos\theta}{m^2} + \frac{\gamma\cos^2\theta}{2m^2} \nonumber \\ & +\frac{1}{m^2}(\gamma\cos\theta+1)\sin\theta\cos\psi\sqrt{\mu-2\bar{\lambda}\bar{p}_\psi} \label{eq:full_ham_nondim} \\
& + \frac{\gamma}{2m^2}\sin^2\theta\cos^2\psi \left(\mu-2\bar{\lambda}\bar{p}_\psi\right) - \frac{\delta\bar{\lambda}}{m^2}\bar{p}_\psi \nonumber
\end{eqnarray}
and
\begin{equation}
\bar{I}_2=\bar{p}_\psi + \frac{\bar{\lambda}\cos\theta}{m^2}-\sqrt{\mu-2\bar{\lambda}\bar{p}_\psi}\left( \bar{p}_{\theta}\sin\psi - \cos\psi \left( \frac{1 - \bar{p}_{\psi}\cos\theta}{\sin\theta} \right) \right).
\label{constraint_nonsim}
\end{equation}

\begin{rem}
$\bar{I}_2$ is not only a first integral of the canonical equations generated by $\bar\mathcal{H}$ for $\gamma=\delta=0$. It is also a first integral for the case $\lambda=0$. This is seen from the more familiar form in \eref{eq:int2} more readily than from the form in \eref{constraint_nonsim} obtained after going through the reduction. In the absence of a magnetic field the parameter $\mu$ is artificial, the result of our choice of $\mathbf{e}_3$.
\end{rem}

Since it seems impossible to use the integral $\bar{I}_2$ to reduce the canonical system further to a single-degree-of-freedom one, homoclinic orbits are not easily obtained in the general case. However, in the non-magnetic case ($\bar\lambda=0$) a further reduction is possible and homoclinic orbits can be obtained explicitly. This is the topic of the next section.

\section{Homoclinic solutions of the extensible rod in zero magnetic field} \label{subsec:extensible}

In the absence of a magnetic field ($\bar\lambda=0$) the force $\boldsymbol{n}$ is a constant vector, by \eref{eq:force}. Provided this vector is not zero, we can choose the fixed frame vector $\boldsymbol{e}_3$ in the direction of $\boldsymbol{n}$ (this gives the usual physical meaning to the Euler angles $\theta$, $\psi$, $\phi$).
According to \eref{eq:magnetic_casimirs1} and \eref{eq:magnetic_casimirs2} the Casimirs satisfy ${2C_{1}=C_{2}^{2} \ne 0}$. Hence $\mu=0$ and the Hamiltonian \eref{eq:full_ham} becomes
\begin{eqnarray}
\bar{\mathcal{H}}(\theta,\bar{p}_\theta,\bar{p}_\psi) = & \frac{1}{2} \bar{p}_{\theta}^{2} + \frac{1}{2} \left( \frac{\bar{p}_{\psi}-\cos\theta}{\sin\theta} \right)^{2} + \frac{\cos\theta}{m^2} + \frac{\gamma\cos^2\theta}{2m^2}.
\label{eq:extensible_ham}
\end{eqnarray}

We are interested in homoclinic orbits, so we assume the rod to be loaded by an end force and end moment applied axially to the rod, which is aligned with $\boldsymbol{e}_3$ as ${t\rightarrow \pm\infty}$. Thus $n_3$ and $m_3$ are the end loads and we have $p_\psi=\mathsf{n}\cdot\mathsf{m}/C_2=I_2/C_2=m_3$, hence $\bar{p}_\psi=1$. Hamilton's equations corresponding to \eref{eq:extensible_ham} in this case read
\begin{equation}
{\dot\theta} = \bar{p}_{\theta} \quad \mbox{and} \quad \dot{\bar{p}}_{\theta} = -\frac{\left(1-\cos\theta\right)^2}{\sin^{3}\theta} + \frac{\left(\gamma\cos\theta+1\right)\sin\theta}{m^{2}}, \label{eq:equivalent}
\end{equation}
where we have used an overdot to denote differentiation with respect to $t$.
This system of equations agrees with that derived in \cite{Stump00} (see also the planar reduction in \cite{Antman75}).

The trivial fixed point $\theta=0$ of \eref{eq:equivalent} corresponds to a straight twisted rod. Non-trivial fixed points solve the cubic
\begin{equation} 
(\gamma\cos\theta+1)(1+\cos\theta)^2=m^2  \nonumber
\end{equation}
and correspond to helical solutions. They exist for
\begin{equation}
0 < m^{2} < 4\left(1+\gamma\right),  \label{eq:buckling_load_2}
\end{equation}
where the upper limit corresponds to the critical load $m_c=2\sqrt{1+\gamma}$ for torsional buckling described by a pitchfork bifurcation \cite{Champneys97a}.

\begin{figure}[tb]
\begin{center}
\subfigure[]{ 
\begin{picture}(0,0)%
\includegraphics{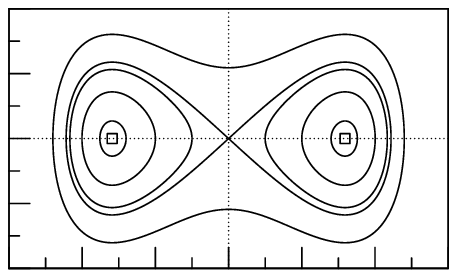}%
\end{picture}%
\begingroup
\setlength{\unitlength}{0.0200bp}%
\begin{picture}(9900,5940)(0,0)%
\put(2475,1650){\makebox(0,0)[r]{\strut{}\small -0.5}}%
\put(2475,2585){\makebox(0,0)[r]{\strut{}\small -0.25}}%
\put(2475,3520){\makebox(0,0)[r]{\strut{}\small 0}}%
\put(2475,4455){\makebox(0,0)[r]{\strut{}\small 0.25}}%
\put(2475,5390){\makebox(0,0)[r]{\strut{}\small 0.5}}%
\put(2750,1100){\makebox(0,0){\strut{}\small -1.5}}%
\put(3804,1100){\makebox(0,0){\strut{}\small -1}}%
\put(4858,1100){\makebox(0,0){\strut{}\small -0.5}}%
\put(5913,1100){\makebox(0,0){\strut{}\small 0}}%
\put(6967,1100){\makebox(0,0){\strut{}\small 0.5}}%
\put(8021,1100){\makebox(0,0){\strut{}\small 1}}%
\put(9075,1100){\makebox(0,0){\strut{}\small 1.5}}%
\put(550,3520){\rotatebox{90}{\makebox(0,0){\strut{}$\bar{p}_{\theta}^{}$}}}%
\put(5912,275){\makebox(0,0){\strut{}$\theta$}}%
\end{picture}%
\endgroup \label{fig:sup_phase} }
\subfigure[]{
\begin{picture}(0,0)%
\includegraphics{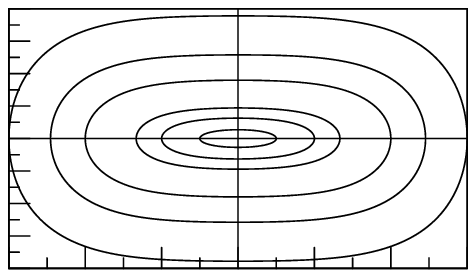}%
\end{picture}%
\begingroup
\setlength{\unitlength}{0.0200bp}%
\begin{picture}(9900,5940)(0,0)%
\put(2200,1650){\makebox(0,0)[r]{\strut{}\small -0.8}}%
\put(2200,2585){\makebox(0,0)[r]{\strut{}\small -0.4}}%
\put(2200,3520){\makebox(0,0)[r]{\strut{}\small 0}}%
\put(2200,4455){\makebox(0,0)[r]{\strut{}\small 0.4}}%
\put(2200,5390){\makebox(0,0)[r]{\strut{}\small 0.8}}%
\put(2475,1100){\makebox(0,0){\strut{}\small -1.5}}%
\put(3575,1100){\makebox(0,0){\strut{}\small -1}}%
\put(4675,1100){\makebox(0,0){\strut{}\small -0.5}}%
\put(5775,1100){\makebox(0,0){\strut{}\small 0}}%
\put(6875,1100){\makebox(0,0){\strut{}\small 0.5}}%
\put(7975,1100){\makebox(0,0){\strut{}\small 1}}%
\put(9075,1100){\makebox(0,0){\strut{}\small 1.5}}%
\put(550,3520){\rotatebox{90}{\makebox(0,0){\strut{}$\bar{p}_{\theta}^{}$}}}%
\put(5775,275){\makebox(0,0){\strut{}$\theta$}}%
\end{picture}%
\endgroup \label{fig:sub_phase} }
\subfigure[]{ 
\begin{picture}(0,0)%
\includegraphics{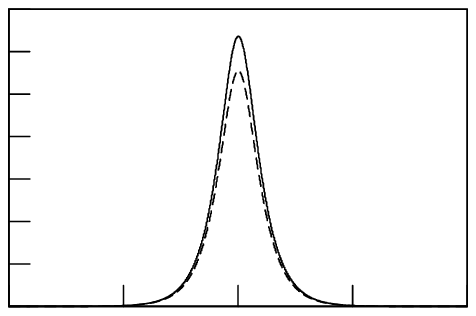}%
\end{picture}%
\begingroup
\setlength{\unitlength}{0.0200bp}%
\begin{picture}(9000,6480)(0,0)%
\put(2200,1650){\makebox(0,0)[r]{\strut{}\small 0}}%
\put(2200,2363){\makebox(0,0)[r]{\strut{}\small 0.2}}%
\put(2200,3077){\makebox(0,0)[r]{\strut{}\small 0.4}}%
\put(2200,3790){\makebox(0,0)[r]{\strut{}\small 0.6}}%
\put(2200,4503){\makebox(0,0)[r]{\strut{}\small 0.8}}%
\put(2200,5217){\makebox(0,0)[r]{\strut{}\small 1}}%
\put(2200,5930){\makebox(0,0)[r]{\strut{}\small 1.2}}%
\put(2475,1100){\makebox(0,0){\strut{}\small -1}}%
\put(4125,1100){\makebox(0,0){\strut{}\small -0.5}}%
\put(5775,1100){\makebox(0,0){\strut{}\small 0}}%
\put(7425,1100){\makebox(0,0){\strut{}\small 0.5}}%
\put(9075,1100){\makebox(0,0){\strut{}\small 1}}%
\put(550,3790){\rotatebox{90}{\makebox(0,0){\strut{}$\theta$}}}%
\put(5325,275){\makebox(0,0){\strut{}$t$}}%
\end{picture}%
\endgroup\label{fig:theta} } 
\subfigure[]{ 
\begin{picture}(0,0)%
\includegraphics{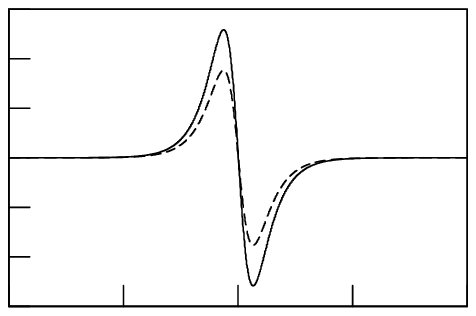}%
\end{picture}%
\begingroup
\setlength{\unitlength}{0.0200bp}%
\begin{picture}(9900,6480)(0,0)%
\put(2200,1650){\makebox(0,0)[r]{\strut{}\small -0.3}}%
\put(2200,2363){\makebox(0,0)[r]{\strut{}\small -0.2}}%
\put(2200,3077){\makebox(0,0)[r]{\strut{}\small -0.1}}%
\put(2200,3790){\makebox(0,0)[r]{\strut{}\small 0}}%
\put(2200,4503){\makebox(0,0)[r]{\strut{}\small 0.1}}%
\put(2200,5217){\makebox(0,0)[r]{\strut{}\small 0.2}}%
\put(2200,5930){\makebox(0,0)[r]{\strut{}\small 0.3}}%
\put(2475,1100){\makebox(0,0){\strut{}\small -1}}%
\put(4125,1100){\makebox(0,0){\strut{}\small -0.5}}%
\put(5775,1100){\makebox(0,0){\strut{}\small 0}}%
\put(7425,1100){\makebox(0,0){\strut{}\small 0.5}}%
\put(9075,1100){\makebox(0,0){\strut{}\small 1}}%
\put(550,3790){\rotatebox{90}{\makebox(0,0){\strut{}$\bar{p}_{\theta}$}}}%
\put(5775,275){\makebox(0,0){\strut{}$t$}}%
\end{picture}%
\endgroup \label{fig:p_theta} } 
\subfigure[]{
\begin{picture}(0,0)%
\includegraphics{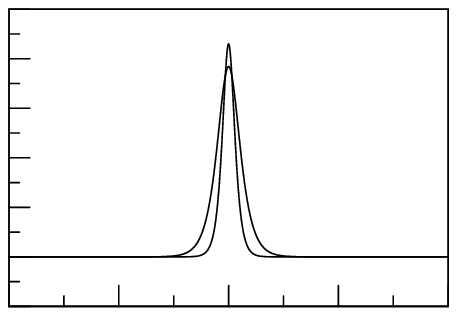}%
\end{picture}%
\begingroup
\setlength{\unitlength}{0.0200bp}%
\begin{picture}(9900,6480)(0,0)%
\put(2475,1650){\makebox(0,0)[r]{\strut{}\small 0.45}}%
\put(2475,2363){\makebox(0,0)[r]{\strut{}\small 0.5}}%
\put(2475,3077){\makebox(0,0)[r]{\strut{}\small 0.55}}%
\put(2475,3790){\makebox(0,0)[r]{\strut{}\small 0.6}}%
\put(2475,4503){\makebox(0,0)[r]{\strut{}\small 0.65}}%
\put(2475,5217){\makebox(0,0)[r]{\strut{}\small 0.7}}%
\put(2475,5930){\makebox(0,0)[r]{\strut{}\small 0.75}}%
\put(2750,1100){\makebox(0,0){\strut{}\small -1}}%
\put(4331,1100){\makebox(0,0){\strut{}\small -0.5}}%
\put(5913,1100){\makebox(0,0){\strut{}\small 0}}%
\put(7494,1100){\makebox(0,0){\strut{}\small 0.5}}%
\put(9075,1100){\makebox(0,0){\strut{}\small 1}}%
\put(550,3790){\rotatebox{90}{\makebox(0,0){\strut{}$\dot\psi$}}}%
\put(5912,275){\makebox(0,0){\strut{}$t$}}%
\end{picture}%
\endgroup \label{fig:psi_prime_ext} }
\end{center}
\caption[]{(a,b) Phase portraits of \eref{eq:eq_oscillator} for $\gamma=1$,
${m =1.7<m_c}$ and $m=2.2>m_c$, respectively.
(c,d,e) $\theta$, $p_{\theta}$ and $\dot\psi$ as a function of scaled arclength $t$ for the homoclinic orbit of the extensible (solid) and inextensible (dashed) rods at $\gamma=1$, ${m=1.7}$.}
\label{fig:angles}
\end{figure}

For parameters satisfying \eref{eq:buckling_load_2} the trivial fixed point is a hyperbolic saddle from which a symmetric pair of homoclinic orbits emanates. To find these we integrate \eref{eq:equivalent} once to obtain
\begin{equation}
\frac{1}{2}\dot\theta^2+V(\theta)=h, \quad\quad
V(\theta)=\frac{1}{2}\frac{1-\cos\theta}{1+\cos\theta} + \frac{\cos\theta}{m^2} + \frac{\gamma}{2m^2}\cos^2\theta
\label{eq:eq_oscillator}
\end{equation}
with $h$ the `energy' level. The corresponding phase portrait is shown in Figure~\ref{fig:sup_phase} (while Figure~\ref{fig:sub_phase} shows a typical phase portrait for $m>m_c$). Setting ${u=\cos\theta}$ and noting that the energy of the homoclinic orbits is 
\begin{eqnarray}
h & = \frac{1}{m^{2}}\left(1+\frac{\gamma}{2}\right) \nonumber
\end{eqnarray}
we can solve \eref{eq:eq_oscillator} for $\dot u$ and integrate to get
\begin{equation}
t = \frac{m}{\sqrt{\gamma}} \int_{u\left(0\right)}^{u\left(t\right)} \frac{ \mathrm{d}u }{ \left(1-u\right) \sqrt{ g\left(u\right) } } \label{eq:integral_g}
\end{equation}
where
\begin{equation}
g\left(u\right) = u^{2} + 2 u\left(1+\frac{1}{\gamma}\right) + 1 + \frac{2}{\gamma} -\frac{m^{2}}{\gamma}. \nonumber
\end{equation}
The quadratic $g\left(u\right)$ has roots
\begin{eqnarray}
u_{\pm} & = -\left(1+\frac{1}{\gamma}\right) \pm \frac{1}{\gamma}\sqrt{ 1 + \gamma m^{2} }. \nonumber
\end{eqnarray}
Substituting the upper and lower bounds for $m$ in \eref{eq:buckling_load_2} into these roots gives
\begin{equation}
-\left(3+\frac{1}{2\gamma}\right) < u_{-} < -\left(1+\frac{2}{\gamma}\right) \quad \mbox{and} \quad -1 < u_{+} < 1. \nonumber
\end{equation}
Thus, the roots of $g$ are always distinct, i.e., ${u_{-} \ne u_{+}}$. The integral~\eref{eq:integral_g} and the limits of integration can be simplified as
\begin{eqnarray}
t & = \frac{m}{\sqrt{\gamma}}\int_{u_{+}}^{u\left(t\right)}\frac{\mathrm{d}u}{\left(1-u\right)\sqrt{\left(u-u_{-}\right)\left( u-u_{+}\right)}}. \label{eq:integration_1}
\end{eqnarray}
The substitution ${u=u_{-} + \left(u_{+}-u_{-}\right)\mathrm{cosh}^{2} z}$ turns the integral into
\begin{equation}
t=\frac{2m}{\left(1-u_{-}\right)\sqrt{\gamma}}
\int_{0}^{z\left(t\right)} \frac{\mathrm{d}z}{1-k\,\mathrm{cosh}^{2}z}
\quad \mbox{with} \quad k=\frac{u_{+}-u_{-}}{1-u_{-}}.
\label{eq:integration_2}
\end{equation}
Note that since $u_{-}<-1$ and ${-1<u_{+}<1}$ we have that ${k>1}$ for all parameter values. Thus the integral~\eref{eq:integration_2} is never singular and may be solved by using the identity
\begin{equation}
\int \!\! \frac{\mathrm{d}z}{1 - k \cosh^{2} z}  = \frac{-2}{\sqrt{k^{2}-1}} \tan^{-1}\left( \sqrt{\frac{k+1}{k-1}}\tanh \frac{z}{2} \right). \nonumber
\end{equation}

Hence, the homoclinic solutions are given by
\begin{eqnarray}
\fl
\cos\theta = u_{-} + \left(u_{+}-u_{-}\right)\cosh^{2} \left(2
\tanh^{-1}\left( \sqrt{\frac{k-1}{k+1}} \tan\left(
\frac{t\left(1-u_{-}\right)\sqrt{\gamma\left(k^{2}-1\right)}}{4m}\right)
\right)\right), \label{eq:theta_homoclinic} \\
\fl
p_{\theta} = {\dot\theta}. \label{eq:p_theta_homoclinic}
\end{eqnarray}
This solution agrees with that derived in \cite[Appendix]{Stump00}, where it is expressed in terms of a natural logarithm rather than hyperbolic functions. In the limit of small extensibility, i.e., ${\gamma\rightarrow 0}$, \eref{eq:theta_homoclinic}
recovers the expression \cite[Eq.~{69}]{Heijden00a}:
\begin{eqnarray}
\theta  = \cos^{-1} \left( u_{0} + \left( 1 - u_{0} \right) \tanh^{2} \left( \frac{ \sqrt{1-u_{0}} }{ m\sqrt{2} } t \right)\right) \quad \mbox{and} \quad p_{\theta} = \dot\theta, \label{eq:Kirchhoff_homoclinic}
\end{eqnarray}
where ${u_{+} \rightarrow u_{0} = m^{2}\slash2 - 1}$, ${u_{-}\rightarrow -\infty}$ and ${k\rightarrow 1}$ as ${\gamma\rightarrow 0}$. Figures~\ref{fig:theta},\ref{fig:p_theta},\ref{fig:psi_prime_ext} compare both homoclinic orbits.

The derivative of the angle $\psi$ is given by
\begin{eqnarray}
{\dot\psi} = \frac{1}{1+\cos\theta}. \label{eq:psi_condition}
\end{eqnarray}
Figures~\ref{fig:theta},\ref{fig:p_theta},\ref{fig:psi_prime_ext} show plots of $\theta$, $p_\theta$ and $\dot\psi$, which we will need in the Mel'nikov analysis.

\section{Mel'nikov theory applied to the magnetically perturbed extensible rod} \label{sec:melnikov}
\par
In order to express the Hamiltonian \eref{eq:full_ham_nondim} in the form \eref{eq:perturbed_hamiltonian} for use in the Mel'nikov analysis we introduce a small parameter $\epsilon$ and write
\begin{eqnarray}
\mu = a \epsilon^{2} &  \quad \mbox{and} \quad \bar\lambda=b\epsilon^{2}, \label{eq:condition_unperturbed_1}
\end{eqnarray}
where $a$ and $b$ are positive and $\mathcal{O}\left(1\right)$, as are the other parameters $m$, $\gamma$ and $\delta$. Then the Hamiltonian takes the form 
\begin{eqnarray}
\mathcal{H}_{}\left(\theta,\psi,\bar{p}_{\theta},\bar{p}_{\psi}\right) & = \mathcal{H}_{0}^{}\left(\theta,\bar{p}_{\theta},\bar{p}_{\psi}\right) + \epsilon \mathcal{H}_{1}^{}\left(\theta,\psi,\bar{p}_\theta,\bar{p}_{\psi}\right)+ \mathcal{O}\left(\epsilon^{2}\right), \nonumber
\end{eqnarray}
where the unperturbed Hamiltonian $\mathcal{H}_{0}$ is given by \eref{eq:extensible_ham} and the first-order perturbation is given by
\begin{eqnarray}
\mathcal{H}_{1}^{}\left( \theta, \psi, \bar{p}_\theta, \bar{p}_{\psi} \right) & = \frac{1}{m^{2}} \left(\gamma\cos\theta+1\right)\sin\theta\cos{\psi}\sqrt{ a - 2 b \bar{p}_{\psi} }. \nonumber  
\end{eqnarray}
For the frequency at the homoclinic orbit \eref{eq:theta_homoclinic} we find
\begin{equation}
\omega_0=\left.\frac{\partial\mathcal{H}_0}{\partial\bar{p}_\psi}\right|_{\mbox{\scriptsize hom}}=\frac{1}{1+\cos\theta}.
\end{equation}
Since $\theta$ in the homoclinic orbit is bounded away from $\pi$, this $\omega_0$ is well-defined and bounded away from zero. The two conditions (i) and (ii) in Section \ref{sec:melnikov_thm} are therefore satisfied and Mel'nikov theory can be applied.

The required partial derivatives are
\begin{eqnarray}
\frac{\partial \mathcal{H}_{0}}{\partial \theta} & = \sin\theta \left( \frac{1}{\left(1+\cos\theta\right)^{2}} - \frac{\left(1+\gamma\cos\theta\right)}{m^{2}} \right), \nonumber \\
\frac{\partial \mathcal{H}_{0}}{\partial \bar{p}_{\theta}} & = \bar{p}_{\theta}, \nonumber \\
\frac{\partial \omega_{0}}{\partial \theta} & = \frac{\sin\theta}{\left(1+\cos\theta\right)^{2}}, \nonumber \\
\frac{\partial \omega_{0}}{\partial \bar{p}_{\theta}} & = 0, \nonumber \\
\frac{\partial \mathcal{H}_{1}}{\partial \theta} & = \frac{1}{m^{2}}\left(\cos\theta+\gamma\cos2\theta\right)\cos\psi\sqrt{a-2 b \bar{p}_{\psi}}, \nonumber \\
\frac{\partial \mathcal{H}_{1}}{\partial \bar{p}_{\theta}} & = 0. \nonumber
\end{eqnarray}
On using \eref{eq:melnikov_bracket} and the identity
\begin{equation}
\left\{ \mathcal{H}_{0}, \frac{\mathcal{H}_{1}}{\omega_{0}} \right\}_{\left(\theta,\bar{p}_\theta\right)}=\frac{1}{\omega_0}
\left\{ \mathcal{H}_{0},\mathcal{H}_{1} \right\}_{\left(\theta,\bar{p}_\theta\right)}
-\frac{\mathcal{H}_{1}}{\omega_0^2}
\left\{ \mathcal{H}_{0},\omega_0 \right\}_{\left(\theta,\bar{p}_\theta\right)},
\end{equation}
and writing ${\psi\left(t\right) = \bar{\psi}\left(t\right) + \psi_{0}}$, with $\bar\psi$ such that $\bar\psi(0)=0$, the Mel'nikov integral is found to be
\begin{eqnarray}
&\fl
\mathcal{M}\left(\psi_{0}^{}\right) = 
\frac{-\sqrt{ a - 2 b}}{m^2}\sin\psi_{0}^{}\int_{-\infty}^{+\infty} \bar{p}_{\theta}\sin\bar{\psi}\left[  \left(1+\cos\theta\right)\left(\cos\theta+\gamma\cos2\theta\right)
\right.
\label{eq:Mel_ext} \\
&\fl
\left.\quad\quad\quad\quad\quad
+\sin^2\theta\left(1+\gamma\cos\theta\right)\right] \mathrm{d}t. \nonumber
\end{eqnarray}
Here we have dropped the $\cos\psi_0$ term which by symmetry does not contribute since $\theta$ is an even function of $t$ while $\bar p_\theta$ and $\bar\psi$ are odd functions of $t$ (cf.~Figure~\ref{fig:angles}).

Generically the Mel'nikov integral will have simple zeroes provided ${a - 2 b \bar{p}_{\psi} \ne 0}$. This condition is no restriction as it corresponds to the non-alignment condition \eref{eq:alignment_condition} which is assumed throughout.
There may be special parameter values for which the Mel'nikov integral is zero.
In these exceptional cases the system will remain nonintegrable but the intersection of the stable and unstable manifolds will be nontransverse. 
\begin{figure}[tb]
\begin{center}
\begin{picture}(0,0)%
\includegraphics[scale=0.8]{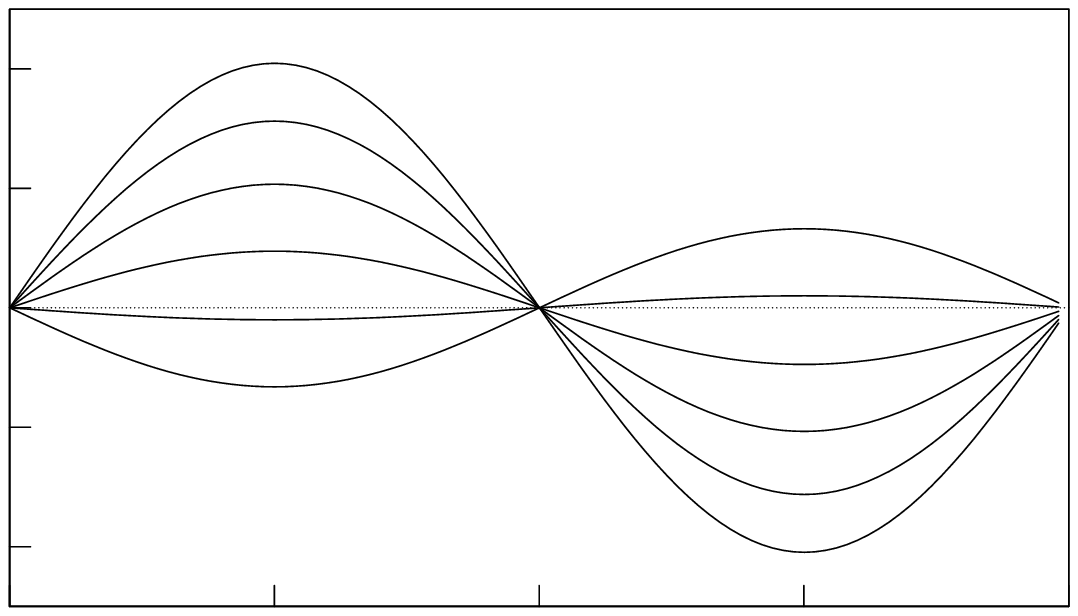}%
\end{picture}%
\begingroup
\setlength{\unitlength}{0.0160bp}%
\begin{picture}(18000,10800)(0,0)%
\put(1650,9390){\makebox(0,0)[r]{\strut{} 2}}%
\put(1650,7670){\makebox(0,0)[r]{\strut{} 1}}%
\put(1650,5950){\makebox(0,0)[r]{\strut{} 0}}%
\put(1650,4230){\makebox(0,0)[r]{\strut{}-1}}%
\put(1650,2510){\makebox(0,0)[r]{\strut{}-2}}%
\put(17175,1100){\makebox(0,0){\strut{}$2\pi$}}%
\put(13362,1100){\makebox(0,0){\strut{}$3\pi\slash{2}$}}%
\put(9550,1100){\makebox(0,0){\strut{}$\pi$}}%
\put(5737,1100){\makebox(0,0){\strut{}${\pi}\slash{2}$}}%
\put(1925,1100){\makebox(0,0){\strut{}0}}%
\put(550,5950){\rotatebox{90}{\makebox(0,0){\strut{}$\mathcal{M}\left(\psi_{0}\right) \slash \sqrt{a-2b}$}}}%
\put(9550,275){\makebox(0,0){\strut{}$\psi_0$}}%
\end{picture}%
\endgroup
\label{fig:Mel_ext_final}
\end{center}%
\caption{Plots of the (normalised) Mel'nikov integral at $m=1.7$ and $\gamma=1$, 1.2, 1.4, 1.6, 1.8 and 2 (from top to bottom at $\psi_0=\pi/2$). The integral is identically zero for a value of $\gamma$ somewhere between 1.8 and 2.}
\label{fig:Mel_ext}
\end{figure}
\par
A plot of the Mel'nikov integral~\eref{eq:Mel_ext} is shown in Figure~\ref{fig:Mel_ext} for various values of $\gamma$ confirming the existence of simple zeroes. We note that for $\gamma$ close to 1.8 the integral is identically zero, indicating a nontransverse intersection of the stable and unstable manifolds.

\section{Numerical results} \label{sec:numerics}
\begin{figure}[tb]
\begin{center}
\subfigure[][$\bar\lambda=0.135$]{
\includegraphics[width=7cm]{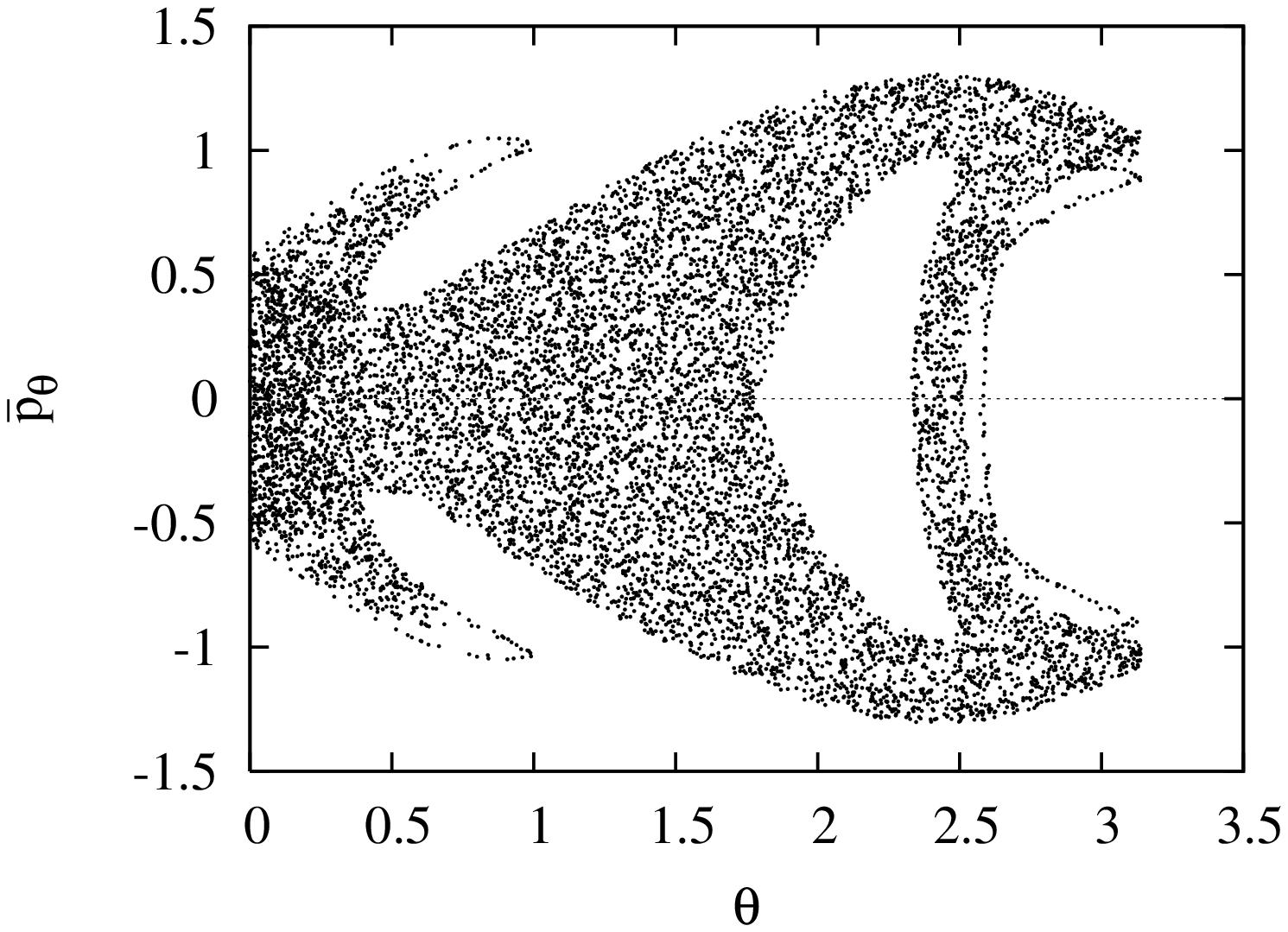}%
\label{fig:poincare1a} }
\subfigure[][$\bar\lambda=0.1575$]{ 
\includegraphics[width=7cm]{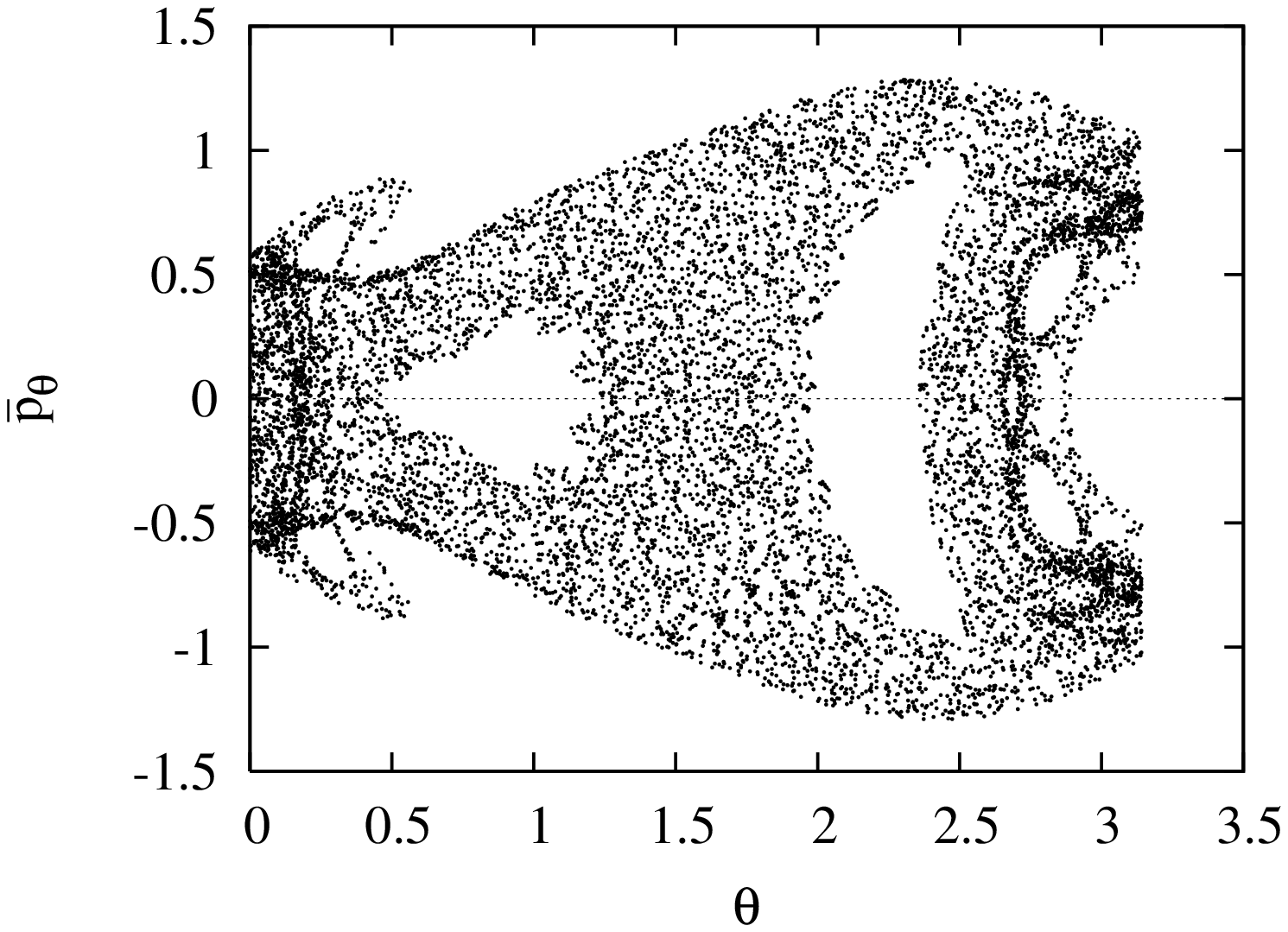}%
\label{fig:poincare1b} }
\subfigure[][$\bar\lambda=0.175$]{ 
\includegraphics[width=7cm]{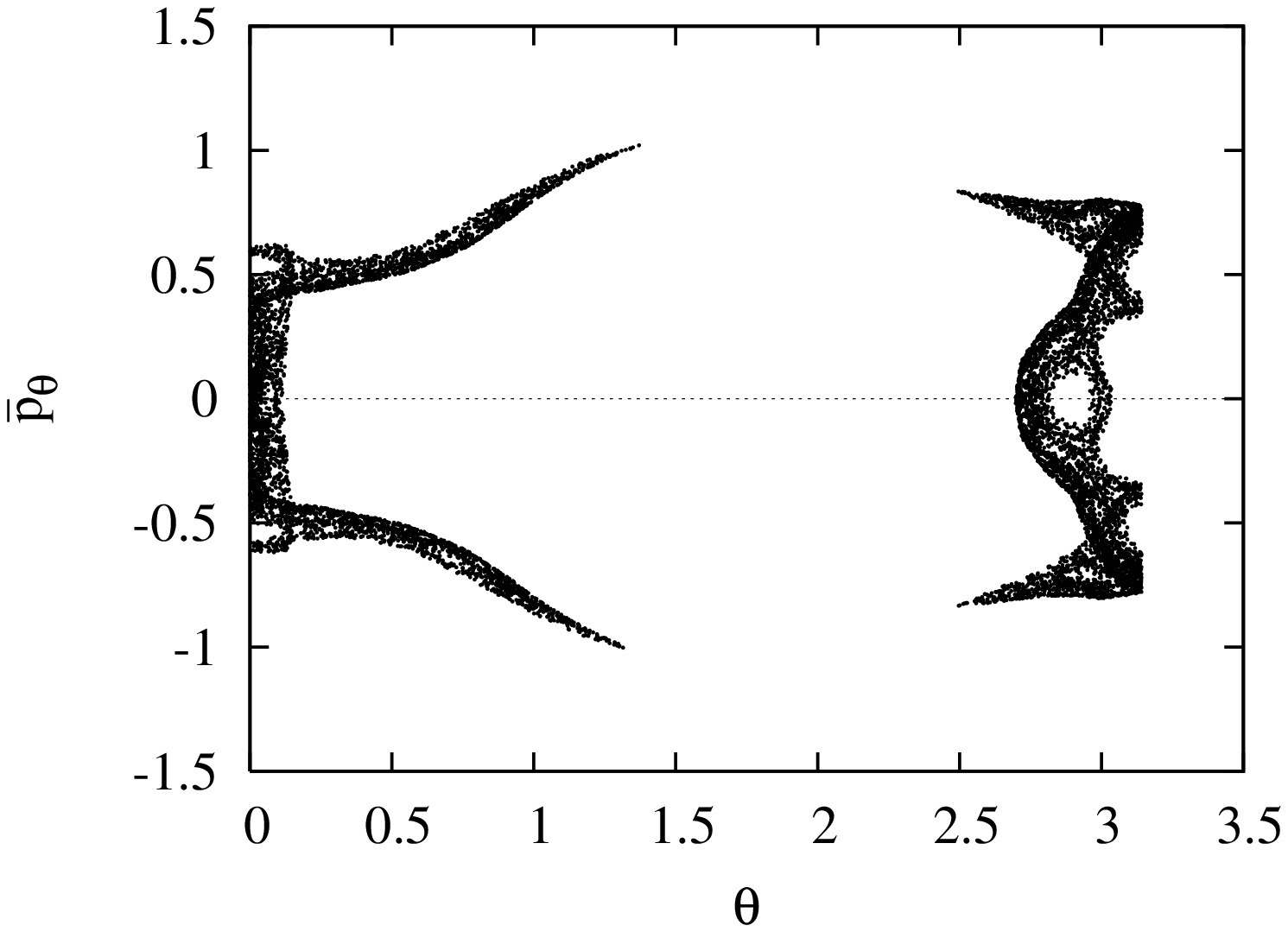}%
\label{fig:poincare1c} }
\subfigure[][$\bar\lambda=0.186$]{ 
\includegraphics[width=7cm]{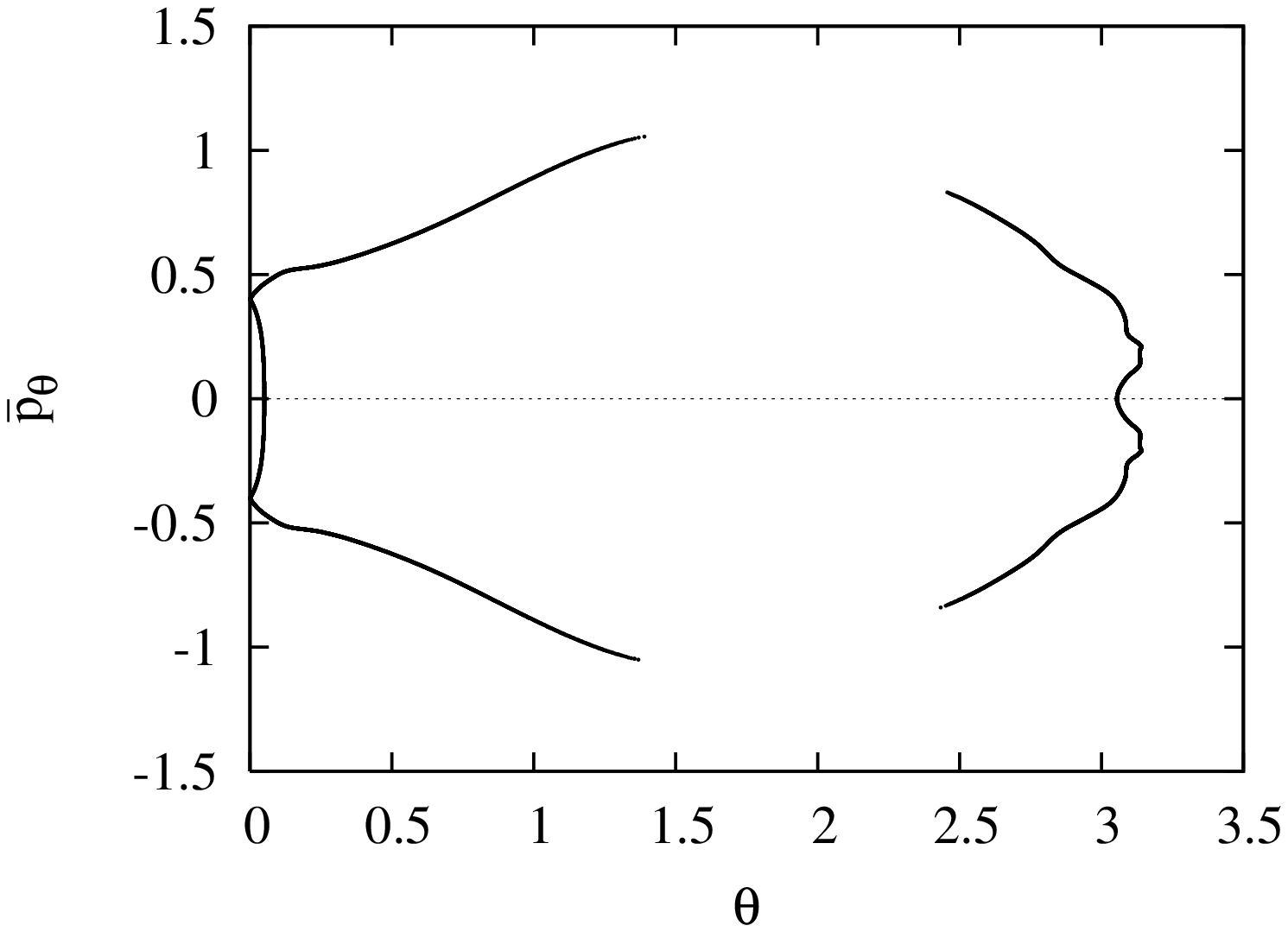}%
\label{fig:poincare1d} }
\end{center}
\caption[Poincar\'e sections on the homoclinic energy level]{Poincar\'e sections for ${\sin\psi=0}$ at energy level ${h=0.9}$ for varying (small) values of $\bar\lambda$. ($m=1.7$, $\mu=0.4$, $\gamma=3$, $\delta=3$.)}
\label{fig:poincare1}
\end{figure}

By Corollary~\eref{cor:chaos} we expect the equations for the extensible magnetic rod to be chaotic, i.e., to contain a horseshoe (at least for small $\bar\lambda$).
To confirm this we present Poincar\'e sections in Figure~\ref{fig:poincare1}. Each panel in the figure shows one orbit with starting values $\theta=0.1$, $\bar{p}_\theta=0.5$, $\psi=0$ at fixed Hamiltonian level $h=0.9$. Solutions were computed using the 8th-order Dormand-Prince code {\tt DOP853} \cite{Hairer08} with relative error tolerance set to $10^{-12}$ and intersections were recorded with plane of section given by $\sin\psi=0$. Plots represent 10000 Poincar\'e iterates taking runs up to $t=84000$ in which the Hamiltonian was found to be preserved to within $3.6\times 10^{-9}$. Clearly visible is the break-up of regular closed orbits into the typical fractal sets of chaotic systems as $\bar\lambda$ is varied.

\begin{figure}[tb]
\begin{center}
\begin{picture}(0,0)%
\includegraphics{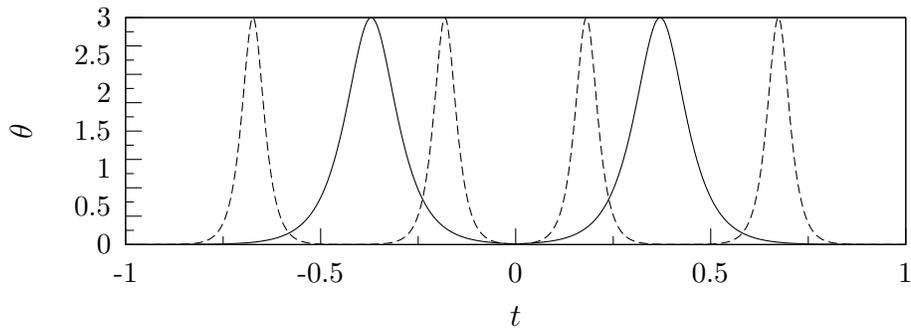}%
\end{picture}%
\begingroup
\setlength{\unitlength}{0.0200bp}%
\begin{picture}(18000,6480)(0,0)%
\put(2200,1650){\makebox(0,0)[r]{\strut{}\small 0}}%
\put(2200,2363){\makebox(0,0)[r]{\strut{}\small 0.5}}%
\put(2200,3077){\makebox(0,0)[r]{\strut{}\small 1}}%
\put(2200,3790){\makebox(0,0)[r]{\strut{}\small 1.5}}%
\put(2200,4503){\makebox(0,0)[r]{\strut{}\small 2}}%
\put(2200,5217){\makebox(0,0)[r]{\strut{}\small 2.5}}%
\put(2200,5930){\makebox(0,0)[r]{\strut{}\small 3}}%
\put(2475,1100){\makebox(0,0){\strut{}\small -1}}%
\put(6150,1100){\makebox(0,0){\strut{}\small -0.5}}%
\put(9825,1100){\makebox(0,0){\strut{}\small 0}}%
\put(13500,1100){\makebox(0,0){\strut{}\small 0.5}}%
\put(17175,1100){\makebox(0,0){\strut{}\small 1}}%
\put(550,3790){\rotatebox{90}{\makebox(0,0){\strut{}$\theta$}}}%
\put(9825,275){\makebox(0,0){\strut{}$t$}}%
\end{picture}%
\endgroup
\end{center}
\caption[]{Two-pulse (solid) and four-pulse (dashed) homoclinic solutions for small $\bar\lambda$ and $\mu$ ($\bar\lambda=0.001$, $\mu=0.002$, ${m=1.7}$, $\gamma=1$, ${\delta=1}$).}
\label{fig:multimodal1}
\end{figure}

Given the existence of a transverse homoclinic orbit in a chaotic system one expects from standard results from dynamical system theory also the presence of  higher-order (multipulse) homoclinic orbits that correspond to solutions that pass near the saddle solution (and the unperturbed homoclinic orbit) multiple times before closing up at the saddle \cite{Belyakov90}. This was found to be the case for anisotropic rods in \cite{Heijden98a}. Multipulse homoclinic orbits for the magnetic rod are displayed in figure~\ref{fig:multimodal1}. They were obtained by means of the shooting method discussed in \cite{Heijden98a}.

\section{Discussion} \label{sec:conc} %
\par 
We have shown that the equilibrium equations for an extensible and shearable conducting rod in a uniform magnetic field are nonintegrable and exhibit chaotic solutions. This is surprising as the effects of extensibility/shearable and magnetic field individually do not lead to a breakdown of integrability of the classical isotropic rod equations. Our system is unlikely to be unique in showing this lack of `additivity' of integrability, but we have not seen it reported of other systems before, physical or otherwise.

\begin{figure}[tb]
\begin{center}
\subfigure[][$\gamma=0.15$]{
\includegraphics[width=7cm]{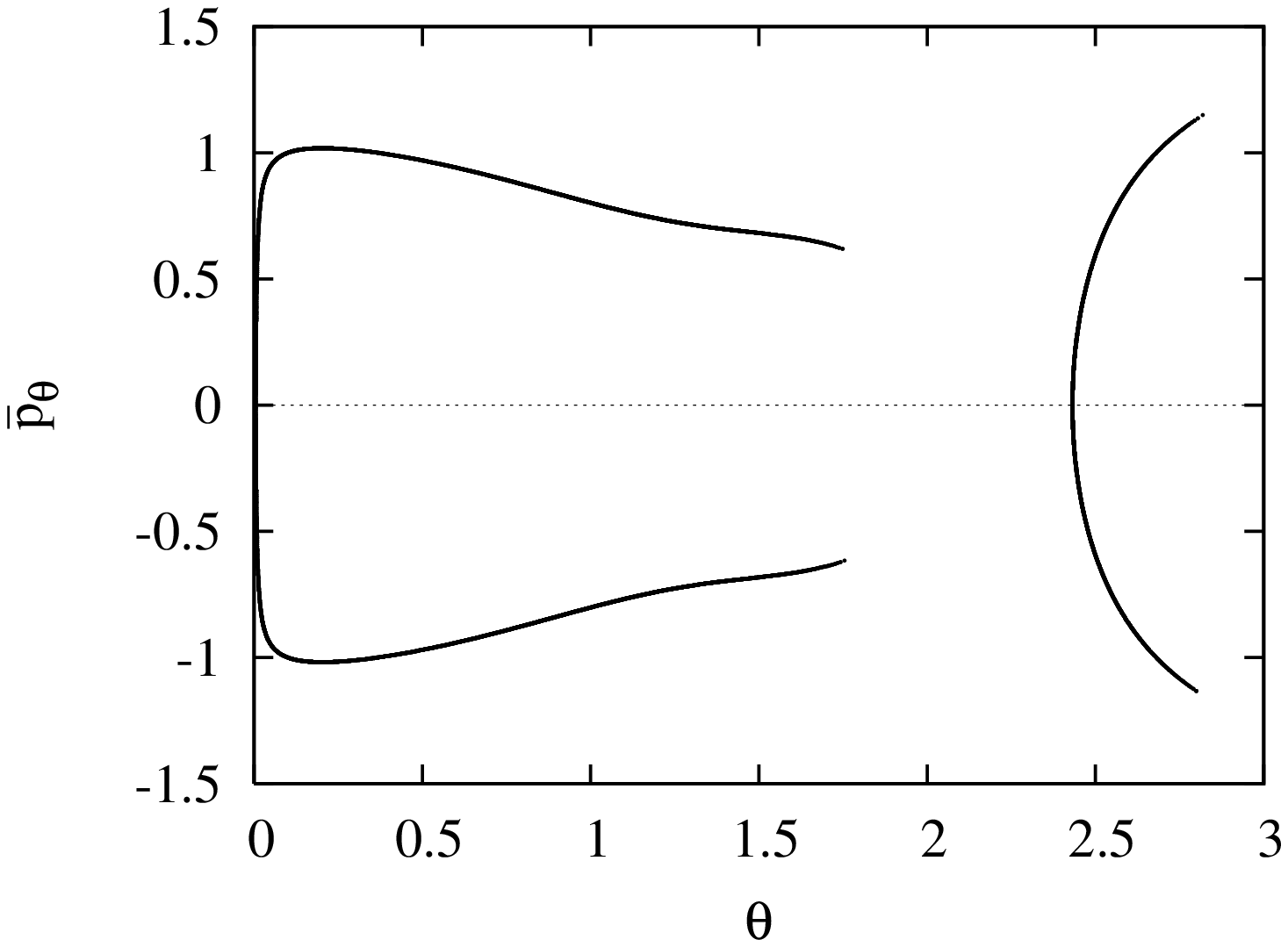}%
\label{fig:poincare2a} }
\subfigure[][$\gamma=0.205$]{ 
\includegraphics[width=7cm]{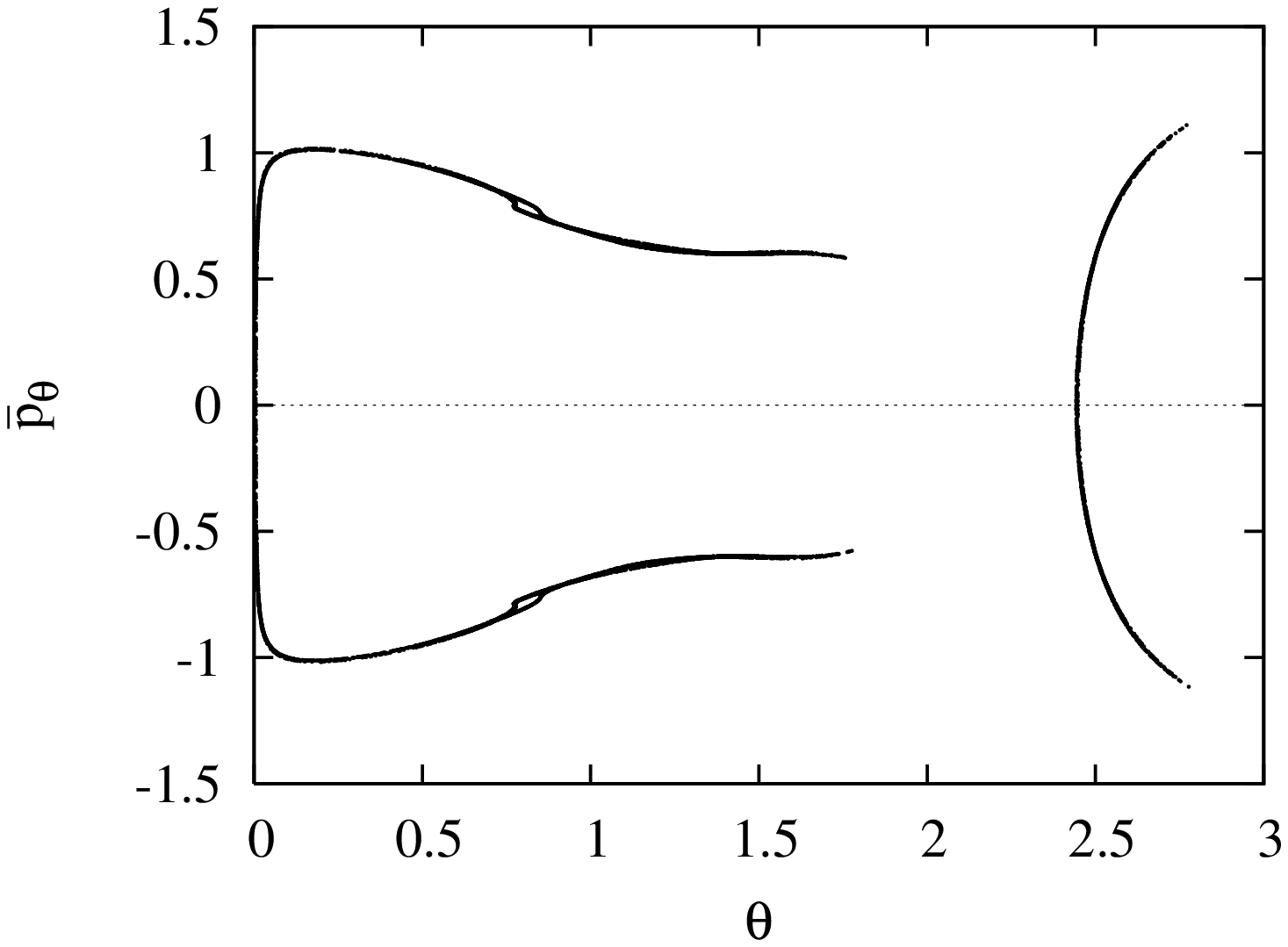}%
\label{fig:poincare2b} }
\subfigure[][$\gamma=0.208$]{ 
\includegraphics[width=7cm]{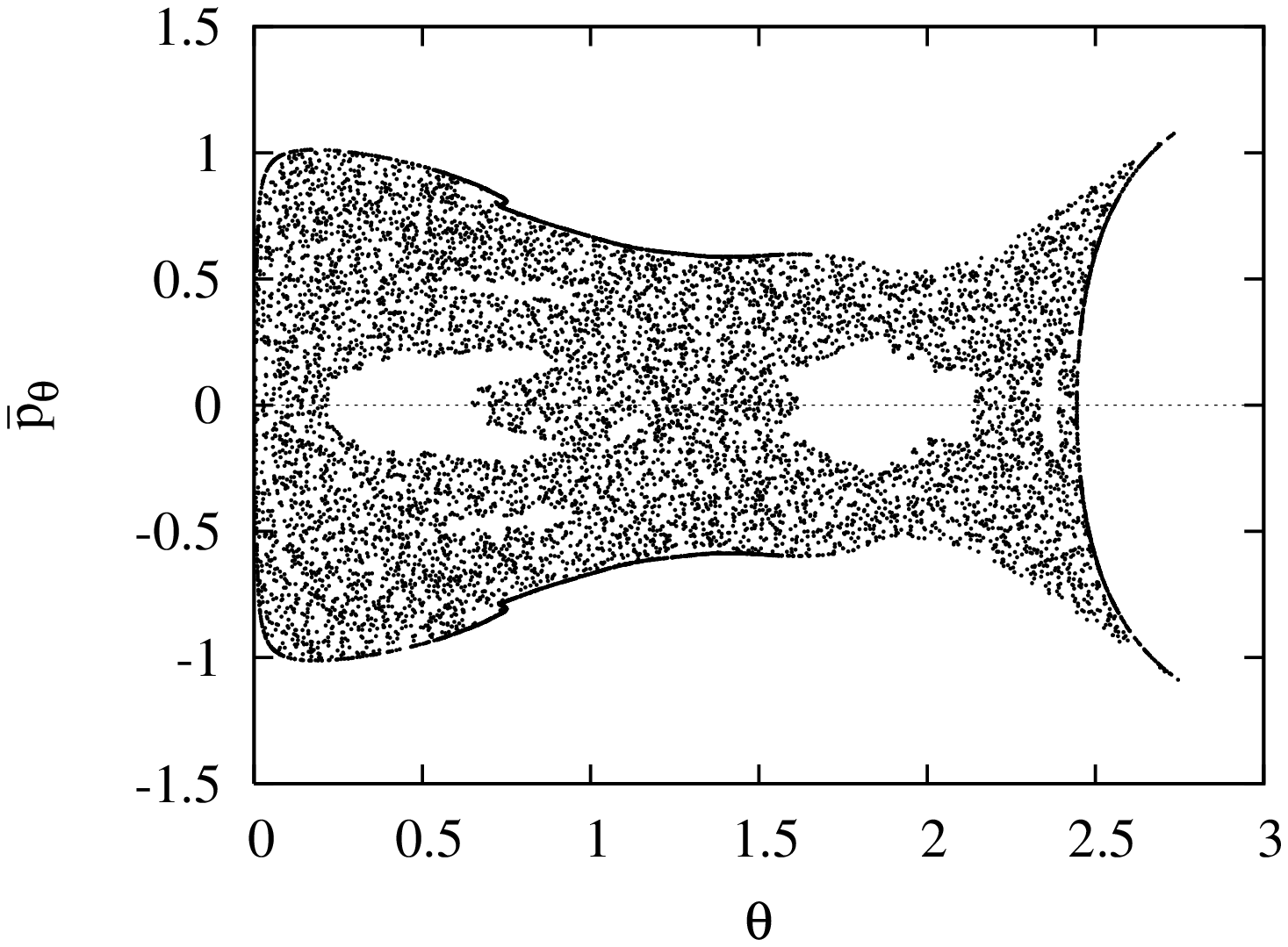}%
\label{fig:poincare2c} }
\subfigure[][$\gamma=0.24$]{ 
\includegraphics[width=7cm]{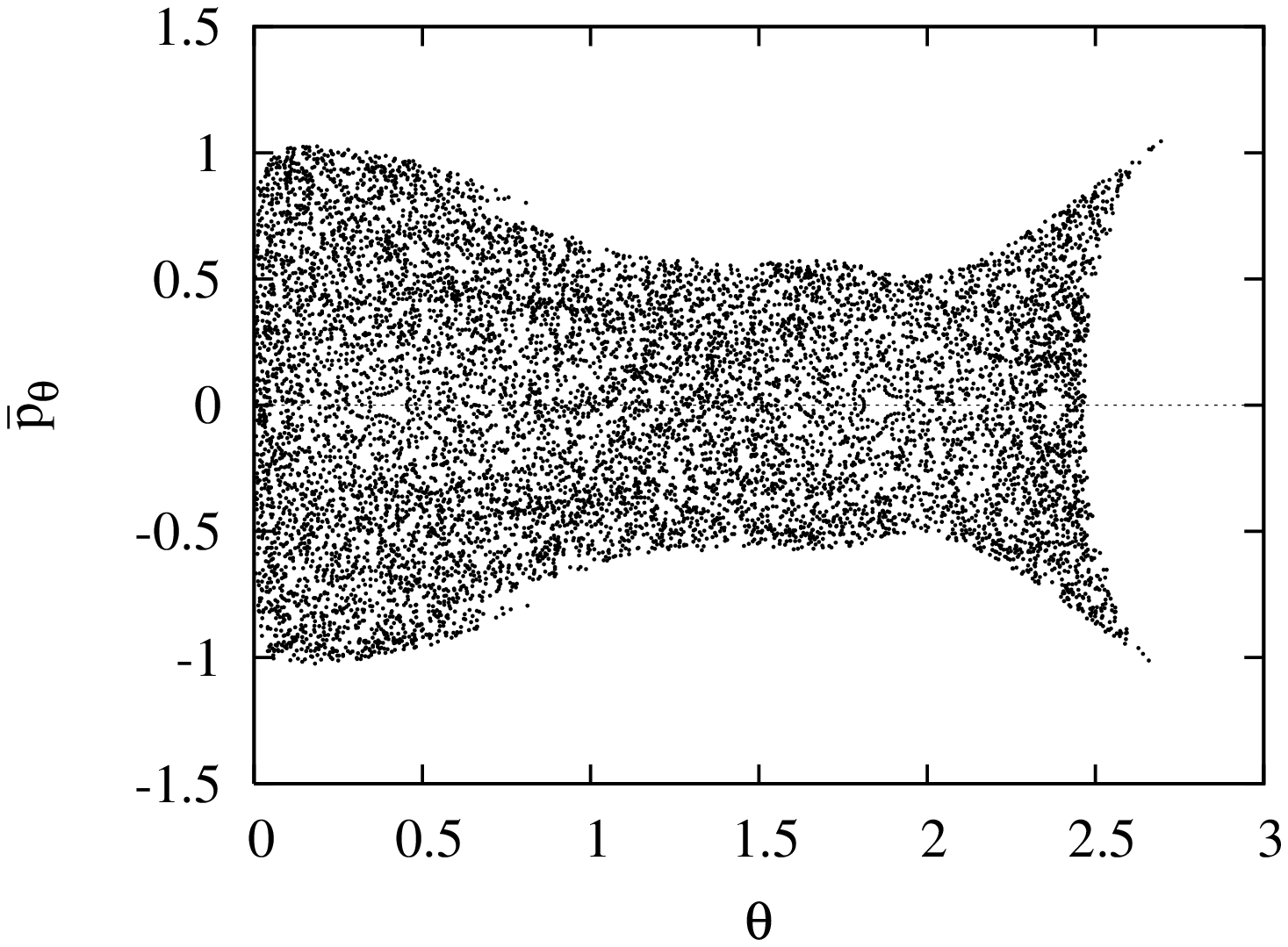}%
\label{fig:poincare2d} }
\end{center}
\caption[Poincar\'e sections on the homoclinic energy level]{Poincar\'e sections for ${\sin\psi=0}$ at energy level ${h=0.8}$ for varying (small) values of $\gamma$. ($m=1.7$, $\delta=0.25$, $\bar\lambda=2$, $\mu=4$.)}
\label{fig:poincare2}
\end{figure}

\begin{figure}[tb]
\begin{center}
\begin{picture}(0,0)%
\includegraphics{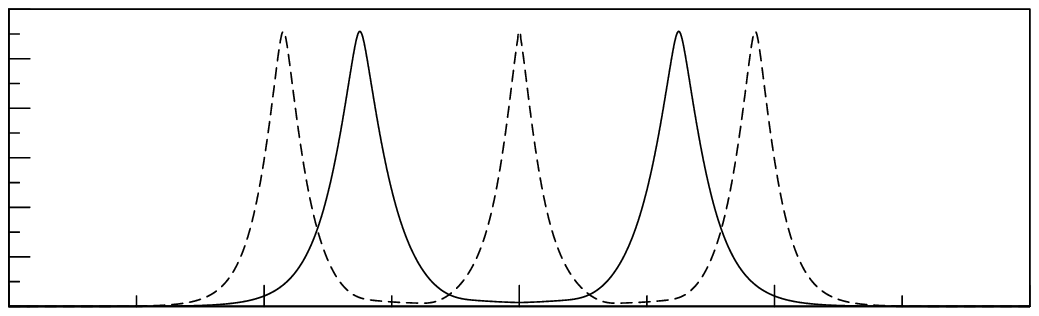}%
\end{picture}%
\begingroup
\setlength{\unitlength}{0.0200bp}%
\begin{picture}(18000,6480)(0,0)%
\put(2200,1650){\makebox(0,0)[r]{\strut{}\small 0}}%
\put(2200,2185){\makebox(0,0)[r]{\strut{}\small 0.2}}%
\put(2200,2720){\makebox(0,0)[r]{\strut{}\small 0.4}}%
\put(2200,3255){\makebox(0,0)[r]{\strut{}\small 0.6}}%
\put(2200,3790){\makebox(0,0)[r]{\strut{}\small 0.8}}%
\put(2200,4325){\makebox(0,0)[r]{\strut{}\small 1}}%
\put(2200,4860){\makebox(0,0)[r]{\strut{}\small 1.2}}%
\put(2200,5395){\makebox(0,0)[r]{\strut{}\small 1.4}}%
\put(2200,5930){\makebox(0,0)[r]{\strut{}\small 1.6}}%
\put(2475,1100){\makebox(0,0){\strut{}\small -1}}%
\put(6150,1100){\makebox(0,0){\strut{}\small -0.5}}%
\put(9825,1100){\makebox(0,0){\strut{}\small 0}}%
\put(13500,1100){\makebox(0,0){\strut{}\small 0.5}}%
\put(17175,1100){\makebox(0,0){\strut{}\small 1}}%
\put(550,3790){\rotatebox{90}{\makebox(0,0){\strut{}$\theta$}}}%
\put(9825,275){\makebox(0,0){\strut{}$t$}}%
\end{picture}%
\endgroup
\end{center}
\caption[]{Two-pulse (solid) and three-pulse (dashed) homoclinic solutions for small
$\gamma$ and $\delta$ ($\gamma=0.01$, $\delta=0.01$, $m=1.7$, ${\bar\lambda=0.5}$, $\mu=1$).}
\label{fig:multimodal2}
\end{figure}

To prove nonintegrability of the system we used Mel'nikov's method with the equations scaled in such a way that the unperturbed rod is extensible/shearable but non-magnetic and the magnetic effect forms the perturbation. This is necessary because the Hamiltonian system for a (inextensible) magnetic rod, although integrable, cannot be explicitly reduced to a one-degree-of-freedom system (at least not globally in terms of Euler angles) and therefore the required explicit expressions of the homoclinic orbit are not available. It is reasonable to expect, however, that integrability is broken more widely in parameter space, including regions of small extensibility ($\gamma\ll 1$) and large magnetic field ($\bar\lambda={\cal O}(1)$). Figure~\ref{fig:poincare2} gives numerical evidence for this in the form of chaotic Poincar\'e plots computed for different parameter values. Each panel in the figure shows one orbit with starting values $\theta=0.1$, $\bar{p}_\theta=1$, $\psi=0$ at fixed Hamiltonian level $h=0.8$. Plots represent 10000 Poincar\'e iterates taking runs up to $t=81000$ in which the Hamiltonian was found to be preserved to within $3.9\times 10^{-9}$. Nonintegrability for these parameters is also confirmed by the multipulse homoclinic orbits for small $\gamma$ and $\delta$ shown in Figure~\ref{fig:multimodal2}.

As the chaotic solutions correspond to spatially complex configurations of the rod, our results may be relevant for electrodynamic space tethers \cite{Cartmell08,Valverde08} and, at an entirely different scale, for beams or ribbons as part of micro- or nanoelectromechanical devices such as sensors, resonators, inductors and actuators \cite{Bell06}. For instance, there is significant interest in nanosprings of small pitch because they allow for large magnetic flux densities \cite{Zhang06}.

For sufficiently slender elastic structures localised (i.e., homoclinic) solutions are the preferred mode of deformation \cite{Heijden00a}. We have presented preliminary numerical results showing that in addition to the transverse homoclinic orbit guaranteed to exist by Mel'nikov theory there exist multipulse homoclinic solutions. It would be interesting to study the bifurcation behaviour of these localised solutions as physical parameters are varied. We intend to take this up in a future publication.

\section*{References} %
\bibliographystyle{unsrt} %
\bibliography{Bibliography,bib_extra} %
\end{document}